\documentclass[10pt, a4paper]{article}

\usepackage{amsmath,amssymb}

\usepackage{graphicx}

\usepackage{color}

\newcommand{\cla}{\textcolor[rgb]{0.77,0.12,0.23}}

\numberwithin{equation}{section}
\newcommand{\C}{\mathbb{C}}
\newcommand{\p}{\partial}
\newcommand{\nn}{\nonumber}
\newcommand{\F}{\mathcal{F}}
\newcommand{\PP}{\mathbb{P}}
\newcommand{\la}{\langle}
\newcommand{\ra}{\rangle}
\newcommand{\al}{\alpha}

\DeclareMathOperator{\res}{Res}
\DeclareMathOperator{\tr}{Tr}

\newtheorem{thm}{Theorem}[section]

\newtheorem{conj}[thm]{Conjecture}

\newtheorem{lem}[thm]{Lemma}

\newtheorem{rmk}[thm]{Remark}

\newenvironment{prf}{\noindent {\it Proof} \ }{\hfill $\Box$}
\newenvironment{prfn}[1]{\noindent {\it Proof of #1} \ }{\hfill $\Box$}

\begin{document}

\title{On the Genus Two Free Energies for Semisimple Frobenius Manifolds}
\author{Boris Dubrovin${}^*$, Si-Qi Liu${}^{**}$, Youjin Zhang${}^{**}$\\
{\small * SISSA, Via Beirut 2-4, 34014 Trieste, Italy}\\
{\small Laboratory of Geometric Methods in Mathematical Physics, Moscow State University}\\
{\small and Steklov Math. Institute, Moscow}\\
{\small ** Department of Mathematical Sciences,
Tsinghua University}\\
{\small Beijing 100084, P. R. China}}
\date{}\maketitle

\begin{abstract}
We represent the genus two free energy of an arbitrary semisimple Frobenius manifold as a sum of contributions associated with dual graphs of certain stable algebraic curves of genus two plus the so-called "genus two G-function". Conjecturally the genus two G-function vanishes for a series of important examples of Frobenius manifolds associated with simple singularities as well as for ${\bf P}^1$-orbifolds with positive Euler characteristics. We explain the reasons for such Conjecture and prove it in certain particular cases.
\end{abstract}

\tableofcontents

\section{Introduction} \label{sec-1}

Let $\left(M, \,\cdot\,, \langle~,~\rangle, e, E\right)$ be a semisimple Frobenius manifold of dimension $n$. With such an object\footnote{It also depends on the choice of a so-called \emph{calibration} of the Frobenius manifold, i.e., the choice of a basis of horizontal sections of the deformed flat connection on $M$. See \cite{DZ} for the details.} one can associate (see \cite{DZ}) a formal series
\begin{equation}\label{gexp}
\F = \sum_{g\geq 0} \epsilon^{2g-2} \F_g({\bf t})
\end{equation}
called \emph{free energy} of the Frobenius manifold (in the framework of the theory of Gromov--Witten invariants its exponential is also called \emph{total descendent potential}). Here
\[
{\bf t} =\left( t^{\alpha, p}\right), \quad \alpha=1, \dots, n, \quad p=0, \, 1, \, 2, \dots
\]
are coordinates on the \emph{large phase space}. They coincide with the time variables of the associated \emph{integrable hierarchy of topological type} (see \cite{DZ}, \cite{htt}). The particular coordinate $x:=t^{1,0}$ plays the role of the spatial variable of the integrable hierarchy. The independent parameter $\epsilon$ in physics literature is called \emph{string coupling constant}. Restricting the free energy onto the \emph{small phase space}
\[
F_g\left( t^{1,0}, \dots, t^{n,0}\right):=\F_g({\bf t})|_{t^{\gamma,p}=0\ (p>0)},
\]
one obtains the generating function of the genus $g$ Gromov--Witten invariants. In particular the function $F_0(t)$, $t=\left( t^{1,0}, \dots, t^{n,0}\right)$, coincides with the potential of the Frobenius manifold.

Denote
\[
v_\alpha({\bf t})=\frac{\p^2 \F_0({\bf t})}{\p t^{1,0} \p t^{\alpha, 0}}, \quad \alpha=1, \dots, n
\]
a particular set of the genus zero correlators.  A remarkable property of the genus expansion \eqref{gexp} says that the higher genus terms can be represented in the form
\begin{equation}\label{gexp1}
\F_g({\bf t})=\hat \F_g\left( v({\bf t}), v_x({\bf t}), \dots, v^{(3g-2)}({\bf t})\right), \quad g\geq 1
\end{equation}
where
\[
v({\bf t})=\left(v^1({\bf t}), \dots, v^n({\bf t})\right)
\]
(raising of the indices is done with the help of the flat metric on $M$). Existence of such a representation first conjectured in \cite{eguchi0} follows from vanishing of certain intersection numbers on the moduli space of stable maps \cite{getzler}; in a more general setting it can also be derived from the bihamiltonian recursion relation of the associated integrable hierarchy of topological  type \cite{DZ}. The functions $\hat \F_g\left( v, v_x, \dots, v^{(3g-2)}\right)$ for $g\geq 2$ depend rationally on the jet variables $v_x$, \dots, $v^{(3g-2)}$ while the expression for $\hat \F_1(v, v_x)$ involves also logarithms (see the formula \eqref{def-f1-zh} below). In sequel the hats will be omitted.

In \cite{DZ} an algorithm was developed for computing $\F_g\left( v, v_x, \dots, v^{(3g-2)}\right)$  with $g\ge 1$ by recursively solving the so-called \emph{loop equation}. In particular 
an explicit formula for the genus two free energy $\F_2=\F_2(v, v_x, v^{(2)}, v^{(3)}, v^{(4)})$ is given for any semisimple Frobenius manifold.
This formula (for convenience of the reader we reproduce it in the Appendix \ref{app2} below) is represented in terms of the Lam\'e coefficients, rotation coefficients and the canonical coordinates of the Frobenius
manifold, which are not easy to compute for a concrete example. In this paper, we show that $\F_2$ can be
rewritten as summation of two parts: the first part is given by correlation functions which is easy to compute in the flat
coordinates, while the second part is still represented in terms of the rotation coefficients and the canonical coordinates,
but it vanishes in many examples such as the simple singularities and the $\PP^1$-orbifolds of $ADE$ type. 

Let us proceed to formulation of the main statements of the present paper.

\begin{thm}  \label{main-thm}
Let $M$ be a semisimple Frobenius manifold of dimension $n$. Denote $\F_2$ the genus two free energy for $M$ given by the formula (3.10.114) from \cite{DZ}, 
see the formula given in Appendix B.
Then
\begin{equation}\label{mthm-zh}
\F_2=\sum_{p=1}^{16} c_p\,Q_p+G^{(2)}(u, u_x, u_{xx}).
\end{equation}
Here each term $Q_p$ corresponds to one of the following sixteen graphs
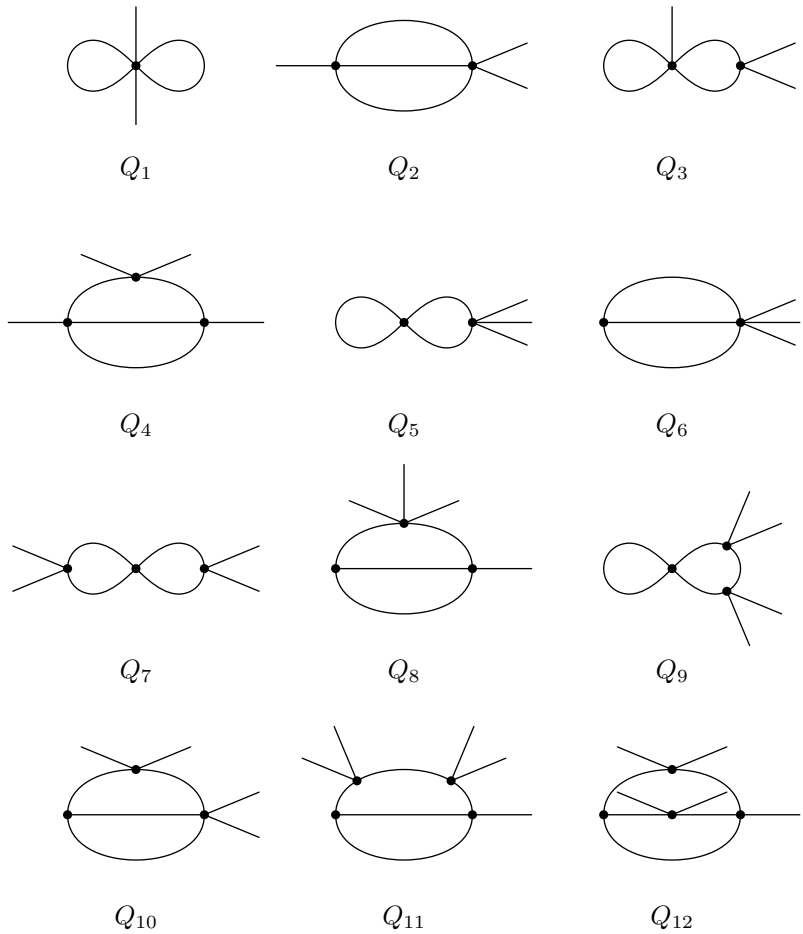
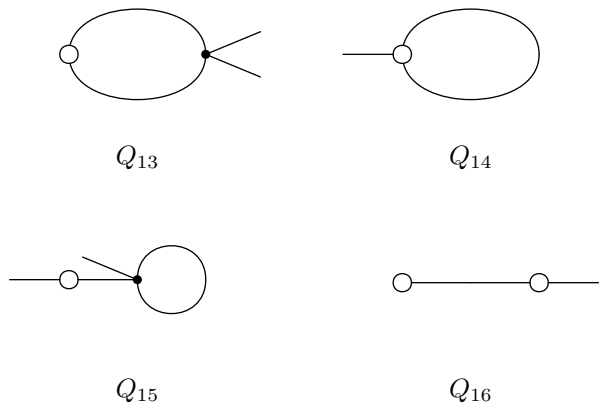
\[
\begin{array}{ccc}
\includegraphics{Q01} & \includegraphics{Q02} &\includegraphics{Q03} \\
Q_1 & Q_2 & Q_3
\end{array}
\]
\[
\begin{array}{ccc}
 \includegraphics{Q04}&\includegraphics{Q05} & \includegraphics{Q06} \\
Q_4 & Q_5 & Q_6
\end{array}
\]
\[
\begin{array}{ccc}
\includegraphics{Q07} & \includegraphics{Q08}& \includegraphics{Q09} \\
Q_7 & Q_8 & Q_9
\end{array}
\]
\[
\begin{array}{ccc}
\includegraphics{Q10} & \includegraphics{Q11} & \includegraphics{Q12} \\
Q_{10}& Q_{11} & Q_{12}
\end{array}
\]
\[
\begin{array}{ccc}
\includegraphics{Q13} & \qquad & \includegraphics{Q14}\\
Q_{13} & & Q_{14}
\end{array}
\]
\[
\begin{array}{ccc}
\includegraphics{Q15} & \qquad & \includegraphics{Q16}\\
Q_{15} & & Q_{16}
\end{array}
\]
The constants $c_p$ read
\begin{align*}
&c_1=0,\quad c_2=-\frac1{960},\quad c_3=\frac1{5760},\quad c_4=\frac1{1152},\\
&c_5=\frac1{2880},\quad c_6=0,\quad c_7=\frac1{1920},\quad c_8=-\frac1{2880},\\
&c_9=-\frac1{1920},\quad c_{10}=\frac1{1920},\quad c_{11}=\frac1{1920},\quad c_{12}=-\frac1{960},\\
&c_{13}=-\frac1{60},\quad c_{14}=\frac1{48},\quad c_{15}=-\frac7{240},\quad c_{16}=\frac7{10}.
\end{align*}
The function $G^{(2)}(u, u_x, u_{xx})$ called the \emph{genus two $G$-function} of the Frobenius manifold. An explicit expression \eqref{g2-zh} of this function in the canonical coordinates $u_1$, \dots, $u_n$ is given in the Appendix \ref{app1}.

\end{thm}

Before formulating the rules for computing the contributions of the sixteen graphs let us explain their realization as \emph{dual graphs} of stable curves of (arithmetic) genus 2. Recall (see, e.g., \cite{manin}) that dual graphs are used to encode a certain class of singular algebraic curves with marked points. Vertices of the graph correspond to the irreducible components of the curve. The genus of the normalization of such a component is called the genus of the vertex. On our sixteen graphs the components of genus zero are shown with bullets; the components of genus 1 are shown with circles. All singularities of a stable curve are at most double points. Points of intersection or self-intersection of these components correspond to the edges of the dual graph while the marked points are associated with the legs. The arithmetic genus of the stable curve is equal to the sum of genera of the vertices plus the first Betti number of the dual graph.

We are now ready to formulate the rules for computing the contribution of a dual graph.
Let $\F_g=\F_g({\bf t})$ be the genus $g=0,1$ free energy of $M$, and
\[\frac{\p}{\p t^{\al,p}},\ \al=1,\dots,n,\ p\ge 0\]
be the tangent vector fields on the big phase space.  We introduce a matrix
\[M_{\alpha\beta}=\frac{\p^3 \F_0}{\p t^{1,0}\,\p t^{\alpha,0}\,\p t^{\beta,0}},\]
and denote $(M^{-1})^{\alpha\beta}$ its inverse. Here and in sequel summation w.r.t. repeated upper and lower indices is assumed. The diagram rules are formulated in the following way:
\begin{itemize}
\item[i)] Bullets ($\bullet$) correspond to $\F_0$;
\item[ii)] Circles ($\circ$) correspond to $\F_1$;
\item[iii)] Edges correspond to $(M^{-1})^{\alpha\alpha'}\frac{\p }{\p t^{\alpha,0}}\otimes\frac{\p }{\p t^{\alpha',0}}$;
\item[iv)] Legs correspond to $\frac{\p}{\p t^{1,0}}$.
\end{itemize}
It is understood that all differential operators corresponding to the edges and legs act first on the vertices $\F_0$ or $\F_1$ and all contractions with the matrix $M^{-1}$ have to be added at the very end.
So, for example,
the terms $Q_1,\ Q_2,\ Q_{15},\ Q_{16}$ are given by
\begin{align}
Q_1=&\frac{\p^6 \F_0}{\p t^{1,0}\,\p t^{1,0}\,\p t^{\alpha,0}\,\p t^{\alpha',0}\,\p t^{\beta,0}\,\p t^{\beta',0}}
\,(M^{-1})^{\alpha\alpha'}\,(M^{-1})^{\beta\beta'},\label{q-1}\\
Q_2=&\frac{\p^4 \F_0}{\p t^{1,0}\,\p t^{\alpha,0}\,\p t^{\beta,0}\,\p t^{\gamma,0}}
\,(M^{-1})^{\alpha\alpha'}\,(M^{-1})^{\beta\beta'}\,(M^{-1})^{\gamma\gamma'}\nn\\
&\times \frac{\p^5 \F_0}{\p t^{1,0}\,\p t^{1,0}\,\p t^{\alpha',0}\,\p t^{\beta',0}\,\p t^{\gamma',0}},\label{q-2}\\
Q_{15}=&\frac{\p^4 \F_0}{\p t^{1,0}\,\p t^{\alpha,0}\,\p t^{\alpha',0}\,\p t^{\beta,0}}(M^{-1})^{\alpha\alpha'}(M^{-1})^{\beta\beta'}\frac{\p^2 \F_1}{\p t^{1,0}\,\p t^{\beta',0}},\label{q-15}\\
Q_{16}=&\frac{\p^2 \F_1}{\p t^{1,0}\,\p t^{\alpha,0}}(M^{-1})^{\alpha\alpha'}\frac{\p \F_1}{\p t^{\alpha',0}}.\label{q-16}
\end{align}
Other $Q_p$'s can be computed in a similar way.

Let us now describe the characteristic properties of the above sixteen graphs that distinguish them from other graphs.
For a graph $Q$, denote $N_v(Q), N_e(Q), N_l(Q)$ the number of its vertices, edges and legs respectively. Let $v_1,\dots, v_m$ with $m=N_v(Q)$ be the vertices of the graph. We also denote $g(v_i), n(v_i)$ the genus and valence of the vertex $v_i$. Finally, $B_1(Q)$ will denote the first Betti number of the graph $Q$.

The above sixteen graphs are selected from the set of connected graphs by requiring that each of these graphs satisfies the following properties:
\begin{enumerate}
\item It is the dual graph of a stable curve of arithmetic genus two.
This property is equivalent to the conditions that the graph is planar, and the valence and genus of its vertices satisfy the constraint $2g(v_i)-2+n(v_i)>0$ and 
\[\sum_{i=1}^m g(v_i)+B_1(Q)=2.\]
\item The number of edges and the number of legs are equal to $N_v(Q)+B_1(Q)-1$.
This property is equivalent to the Euler formula for the graph 
\[N_e(Q)-N_v(Q)+1=B_1(Q)\]
 together with the condition that the function associated to $Q$, defined as above, must have degree two with respect to the jet variables $\p_x^p v^\al$, i.e.
\[\sum_{i=1}^m (2 g(v_i)-2+n(v_i))-N_e(Q)=2\,.\]
Recall that, according to  \cite{D1, DZ-cmp} such a function can be represented as a rational function of the jet variables $\p_x^p v^\al$, $ p\ge 1$, and its degree is defined by assigning degree $p$ to $\p_x^p v^\al$, $\al=1,\dots,n$. We also note that
\[\sum_{i=1}^m 
n(v_i)=2 N_e(Q)+N_l(Q).\]
\item Cutting of an edge connecting two genus zero vertices does not destroy the connectivity of the graph.
A graph with this property is called to be  \emph{one-particle irreducible (1PI)} in physics literature.
\item There is at most one vertex with valence $n(v_i)= 3-2g(v_i)$ in the graph. Moreover,
if the graph contains only one genus one vertex, then the valence of each of its vertices $v_i$ satisfies
$n(v_i)>3-2g(v_i)$.
\end{enumerate} 

\begin{rmk}
If a graph $\tilde{Q}$ is obtained from a graph $Q$ by adding a genus zero vertex with a leg in the middle of an edge of  $Q$, then the functions associated to $\tilde{Q}$ and $Q$ are equal. This follows immediately from the above definitions. So we will view the new graph  $\tilde{Q}$ as same as the old one $Q$.
\end{rmk}

The main point of the decomposition \eqref{mthm-zh} of the genus two free energy into a sum of 16+1 terms is the following

\begin{lem} The restrictions of the terms $Q_1$, \dots, $Q_{16}$ onto the small phase space vanish.
\end{lem}

The proof of the lemma easily follows from the above explicit expressions, the following rules of restricting the jets
\[
{v_x}|_{\mbox{\small small phase space}}=e, \quad {v^{(k)}}|_{\mbox{\small small phase space}}=0
\mbox{ for } k\geq 2,
\]
and the identity
\[\p_e G=0\]
(see details in \cite{DZ}). Here $e$ is the unit of the Frobenius manifold and $G$
is the G-function of the Frobenius manifold that appears in \eqref{def-f1-zh} below.

\medskip

Thus, the part of the free energy ``responsible" for the would-be genus two Gromov--Witten invariants (i.e., with no descendents) is entirely contained in our genus two $G$-function.

Another important feature of the genus two $G$-function can be observed in the analysis of important examples coming from singularity theory and orbifold Gromov--Witten invariants. In the present paper we consider the two classes of examples: first, the case of simple singularities and, the second, the Gromov--Witten invariants of $\PP^1$-orbifolds with positive Euler characteristic. Both classes of examples are associated with Dynkin diagrams of $ADE$ type. The connection of the simple singularities with the $ADE$ Weyl groups is well known. The Frobenius structure on the base of universal unfolding in this case can be constructed with the help of K. Saito theory of primitive forms \cite{saito}. The integrable hierarchies of topological type coincide with the Drinfeld--Sokolov $ADE$ hierarchies \cite{gm, dlz, wu}. 
The associated cohomological field theory was constructed in \cite{Witten4, for-1, for, for-3, fjr-d4, fsz}.

The case of $\PP^1$-orbifolds is relatively more recent. In this case one deals with the $\PP^1$-orbifolds of positive Euler characteristic. Hence there are at most three orbifold points with multiplicities $p$, $q$ and $r$. These positive integers must satisfy
\[
\frac1{p}+\frac1{q}+\frac1{r}>1.
\]
Such an inequality has only finite number of solutions given in the following table
\[
\begin{array}{cc} (p,q,r) & \hskip 1cm  \mbox{Dynkin diagram}\\
\\
(p,q,1) & \hskip 1cm \tilde A_{p,q}\\
(2,2,r) & \hskip 1cm \tilde D_{r+2}\\
(2,3,r) & \hskip 1cm \tilde E_{r+3}
\end{array}
\]
The second column of this table refers to the so-called extended affine Weyl groups of $ADE$ type. The Frobenius manifolds in these cases were constructed in \cite{dz-comp}. The construction depends on the choice of a vertex of the Dynkin diagram. A connection between these Frobenius manifolds and the orbifold quantum cohomology of the $\PP^1$-orbifolds was discovered in \cite{milanov} for the $\tilde A_{p,q}$ case and in \cite{rossi} for other Dynkin diagrams. Also an important connection of these Frobenius manifolds with Frobenius structures on the spaces of the so-called tri-polynomials (see below) was established in \cite{rossi} (the role of tri-polynomials in the homological mirror symmetry was unraveled in \cite{taka}).

The main conjecture of the present paper is the following 

\begin{conj}\label{main-conj1}
If $M$ is the Frobenius manifold obtained from the genus zero Fan--Jarvis--Ruan--Witten (FJRW) invariants theory for ADE singularities,
or the genus zero Gromov--Witten invariants theory for $\PP^1$-orbifolds of ADE type, then
\begin{equation}
G^{(2)}(u, u_x, u_{xx})=0.
\end{equation}
\end{conj}

\begin{rmk}
In FJRW theory there is also involved a symmetry group $G$. We assume that the singularities and their symmetry
groups are chosen so that the corresponding Frobenius manifolds coincide with the usual ones constructed
from the singularities of the same type \cite{D1}. In particular, when the singularities are of $A$ and $E$ type, or $D$
type with even Milnor number, the group $G$ can be chosen as the minimal one $\langle J \rangle$. For the singularities of $D$ type with odd Milnor number,
one need to start with the mirror of $D_n$, i.e. $D^T_n=x^{n-1}y+y^2$, and choose the group $G$ to be
the maximal one $G_{max}$. The reason is that the FJRW theory is a A-model theory, while the 
construction given in \cite{D1} from singularities to Frobenius manifolds is on the B-side, so there is mirror symmetry phenomenons between them.
For more details, see \cite{for, fjr-d4}.
\end{rmk}

The main conjecture can also be formulated in the following way.

\begin{conj}\label{main-conj}
If $M$ is a Frobenius manifold associated to an ADE singularity or an extended affine Weyl groups of ADE type, then
\begin{equation}
G^{(2)}(u, u_x, u_{xx})=0.
\end{equation}
\end{conj}

The validity of this Conjecture has been verified in many examples; the main goal of the present paper is to explain the tools relevant for such a verification.

\begin{rmk} In \cite{eguchi1, eguchi2} formulae for the genus two free energies for the Frobenius manifolds
associated to $A_2$ singularity and to the extended affine Weyl group $\tilde{W}(A_1)$ are given. They have the following graph representations respectively
\begin{align}
&\F_2=\frac1{1152}\,Q_1-\frac1{360}\,Q_2-\frac1{1152}\,Q_3+\frac1{360}\,Q_4,\nn\\
&\F_2=\frac1{1152}\,Q_1-\frac1{360}\,Q_2-\frac1{1152}\,Q_3+\frac1{360}\,Q_4-\frac1{480}\,W_1+
\frac{7}{5760}\,W_2+\frac{11}{5760}\,W_3.\nn
\end{align}
Here $W_1, W_2, W_3$ are the graphs
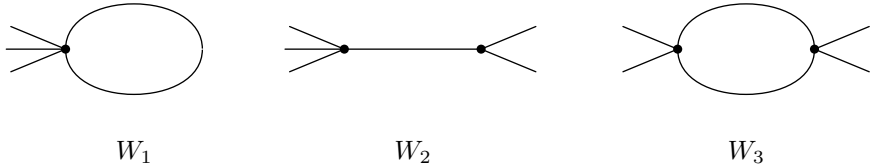
\[
\begin{array}{ccccc}
\includegraphics{Q24} & \hskip -2cm  & \includegraphics{Q25} & \qquad  & \includegraphics{Q26}\\
W_{1} & & W_{2} && W_3
\end{array}
\]
\end{rmk}

When computing the coefficients $c_p$ for our examples, we find the following interesting identity.
\begin{thm} \label{thm-relation}
If $M$ is the Frobenius manifold obtained from the genus zero FJRW invariants theory for $ADE$ singularities,
or the genus zero Gromov--Witten invariants theory for $\PP^1$-orbifolds of AD type, then
\begin{align}
&(Q_1-Q_6)+2(Q_7-Q_5)+3(Q_8- Q_2)\nn\\
&\qquad\qquad+4(Q_9- Q_3)+6 (Q_4+Q_{10}-Q_{11}- Q_{12})=0. \label{relation}
\end{align} 
\end{thm}
The identity \eqref{relation} holds also true for an arbitrary two-dimensional semisimple Frobenius manifold (i.e., for a topological field theory with two primary fields in the terminology of \cite{eguchi2}) as well as for the three-dimensional Frobenius manifolds on the orbit spaces of Coxeter groups of type $B_3$ or $H_3$.  It is interesting to find out in general the necessary and sufficient conditions for validity of this identity.

The paper is organized as follows. In Section \ref{sec-21} we recall first some basic properties of semisimple Frobenius 
manifolds and their genus zero, one and two free energies. Then we give a proof of Theorem \ref{main-thm}.
In Section \ref{sec-22} we prove the Theorem \ref{thm-relation}. In Section \ref{sec-23} and \ref{sec-24}
we give some general formulae in order to calculate the rotation coefficients for Frobenius manifolds  arising in singularity theory. In Section \ref{sec-3} we present more explicit formulae for the rotation coefficients case by case for simple singularities of $ADE$ type and for $\mathbb{P}^1$ orbifolds of $A$ and $D$ type, and provide evidences to support the validity of the conjectures. In the Appendices we give the formulae for the function $G^{(2)}(u,u_x, u_{xx})$ that appear in \eqref{mthm-zh} and for the genus two
free energy of semisimple Frobenius manifolds that is given in \cite{DZ}. 

\section{General Results} \label{sec-2}

\subsection{Proof of Theorem \ref{main-thm}} \label{sec-21}

For a semisimple Frobenius manifold $M^n$, we denote $v^1,\dots, v^n$ the flat coordinates, $\langle\, ,\,\rangle$ its flat metric,
\[
\left\langle \frac{\p}{\p v^\alpha}, \frac{\p}{\p v^\beta}\right\rangle=\eta_{\alpha\beta},\quad (\eta^{\alpha\beta})=(\eta_{\alpha\beta})^{-1},
\] 
and $F(v)=F(v^1,\dots, v^n)$
its potential. The canonical coordinates $u_1,\dots, u_n$ are defined so that
the multiplication table defined on the tangent spaces 
is given by
\[\frac{\p}{\p u_i}\cdot \frac{\p}{\p u_j}=\delta_{ij} \frac{\p}{\p u_i}.\]
In the canonical coordinates the flat metric takes the diagonal form 
\[ \sum_{\al, \beta}\eta_{\al\beta} d v^\al d v^\beta=\sum_{i=1}^n \eta_{ii}(u) d u_i^2.\]
Denote 
\[h_i=h_i(u)=\sqrt{\eta_{ii}},\quad i=1,\dots,n\]
the Lam\'e coefficients of the diagonal metric for some choice of the signs of the square roots.
Define the rotation coefficients $\gamma_{ij}=\gamma_{ji}$ by
\[\gamma_{ij}=\frac1{h_i}\frac{\p h_j}{\p u_i}\quad \mbox{for}\quad i\neq j, \quad \gamma_{ii}=0.\]
The nonzero Christoffel symbols of the Levi-Civita connection for the flat metric in the canonical coordinates
are written in the following table
\begin{equation}\label{gamma-zh}
\Gamma_{ij}^k=\left\{\begin{array}{rl} -\sum_{l=1}^n \gamma_{il} \frac{h_l}{h_i},\quad &i=j=k;\\
\gamma_{ij} \frac{h_j}{h_i},\quad &k=i\ne j;\\
\gamma_{ij} \frac{h_i}{h_j},\quad &k=j\ne i;\\
-\gamma_{ik} \frac{h_i}{h_k},\quad &k\ne i= j.
\end{array}\right.
\end{equation}

The canonical and the flat coordinates of the Frobenius manifold are related by the following equations
\begin{equation}\label{chr-zh}
\frac{\p^2 v^\al}{\p u_i \p u_j}=\sum_{k=1}^n \Gamma_{ij}^k \frac{\p v^\al}{\p u_k}.
\end{equation}

We denote
\[\psi_i^\al(u)=\frac1{h_i(u)}\frac{\p v^\al(u)}{\p u_i},\quad
\psi_{i\al}=\eta_{\al\beta} \psi^\beta_i,\]
where summation w.r.t. repeated upper and lower Greek indices is 
assumed. Assuming that the unit vector field of the Frobenius manifold is $e=\frac{\p}{\p v^1}$,
then 
\begin{equation}\label{psi-zh}
\psi_{i1}=h_i
\end{equation}
and 
\begin{equation}
c_{\al\beta\gamma}=\frac{\p^3 F(v)}{\p v^\al\p v^\beta\p v^\gamma}
=\sum_{i=1}^n \frac{\psi_{i\al}\psi_{i\beta}\psi_{i\gamma}}{\psi_{i1}}.
\end{equation} 
The following formulae \cite{D1} will be used below to represent the correlation functions in terms of the canonical coordinates 
\begin{align}
&\frac{\p v^\al}{\p u_i}=\psi_{i1}\psi_i^\al,\quad \frac{\p u_i}{\p v^\al}=\frac{\psi_{i\al}}{\psi_{i1}};\nn\\
&\frac{\p\psi_{i\al}}{\p u_k}=\gamma_{ik}\,\psi_{k\al},\quad i\ne k,
\quad \frac{\p \psi_{i\al}}{\p u_i}=-\sum_{k=1}^n  \gamma_{ik}\psi_{k\al};\label{dgamma-zh}\\
&\frac{\p \gamma_{ij}}{\p u_k}= \gamma_{ik} \gamma_{kj},\  i, j, k \mbox{ distinct}, \quad\frac{\p \gamma_{ij}}{\p u_i}=
\frac{\sum_{k=1}^n (u_j-u_k) \gamma_{ik}\gamma_{kj}- \gamma_{ij}}{u_i-u_j}.\nn
\end{align}

The principal hierarchy associated to the Frobenius manifold  is a hierarchy of
integrable Hamiltonian systems of hydrodynamic type
\[\frac{\p v^\al}{\p t^{\beta,q}}=\eta^{\al\gamma}\frac{\p}{\p x}\left( \frac{\p\theta_{\beta,q+1}}{\p v^\gamma}\right),
\quad \al, \beta=1,\dots, n,\ q\ge 0.\]
Here $\theta_\al(v;z)=\sum_{p\ge 0} \theta_{\al,p}(v) z^p,\ \al=1,\dots,n$ are related 
to the flat coordinates of the deformed flat connection of the Frobenius manifold. They 
satisfy the conditions
\begin{align*}
&\theta_{\al}(v;0)=\eta_{\al\gamma} v^\gamma,\\
&\langle\nabla\theta_\al(v,-z), \nabla\theta_\al(v,z)\rangle
=\eta_{\al\beta},\\
&\p_\al\p_\beta\theta_\gamma(v;z)=z c^\xi_{\al\beta}\p_\xi\theta_\gamma(v;z),\\
&E\left(\p_\beta\theta_{\alpha,p}(v)\right)=p\, \p_\beta\theta_{\alpha,p}(v)+
\hat{\mu}^\gamma_\alpha \p_\beta\theta_{\gamma,p}(v)+\hat{\mu}^\gamma_\beta \p_\gamma\theta_{\alpha,p}(v)\\
&\hskip 2.3cm  +\left(R_0\right)^\gamma_\beta \p_\gamma\theta_{\alpha,p}(v)
+\sum_{k=0}^p\p_\beta\theta_{\gamma,p-k}(v)\,\left(R_k\right)^\gamma_\alpha,
\end{align*}
where  $E$ is the Euler vector field of the Frobenius manifold which has the 
following representations in the flat coordinates and in the canonical coordinates
respectively
\[ 
E=\sum_{\alpha=1}^n E^\alpha(v) \frac{\p}{\p v^\alpha}=\sum_{i=1}^n u_i \frac{\p}{\p u_i},
\] 
and $\hat{\mu}$ and $R_0$ are the semisimple and nilpotent parts of the antisymmetric constant matrix ${\cal{V}}=({\cal{V}}^\alpha_\beta)$ with 
\[{\cal{V}}^\alpha_\beta=\frac{2-d}2\, \delta^\alpha_\beta - \frac{\p E^\alpha(v)}{\p v^\beta}.\]
The constant matrices $R_0, R_1, \dots R_m$ ($m$ is a certain integer depending on the Frobenius manifold)
form part of the monodromy data of the Frobenius manifold at $z=0$ (see \cite{D1} for detail), they have the properties
\[ (R_k)_\alpha^\gamma\, \eta_{\gamma\beta}=(-1)^{k+1} (R_k)_\beta^\gamma \,\eta_{\gamma\alpha},\quad [\hat{\mu}, R_k]=k R_k,\quad k=0,1,\dots,m.\]
The potential $F(v)$ can be chosen in such a way that the functions $\theta_{\al,1}(v)$
have the expression 
\[\theta_{\al,1}(v)=\p_\al F(v), \quad \p_\al=\frac{\p}{\p v^\al}.\]
Thus the first set of equations of the  principal hierarchy reads
\[\frac{\p v^\al}{\p t^{\beta,0}}=\eta^{\al\xi}c_{\xi\beta\gamma}(v) v^\gamma_x
\quad \textrm{with}\ \ \frac{\p v^\al}{\p t^{1,0}}=v^\al_x,\quad \al,\beta=1,\dots,n.\]
By using the above formulae we get the following formula for solutions of the principal hierarchy:
\begin{equation}\label{du-zh-1}
\frac{\p u_i}{\p t^{\al,0}}\frac{\p v^\al}{\p u_j}
=\frac{\p u_i}{\p t^{\al,0}}\psi^\al_i\psi_{i1}=\left\{\begin{array}{ll} u_{i,x},\quad 
&\textrm{if}\ \ i=j,\\
0,\quad &\textrm{if}\ \ i\ne j.\end{array}\right.
\end{equation}
Moreover, for higher jets $u_i^{(p)}=\p_x^p u_i$ denote 
\begin{equation}\label{du-zh-2}
U^{i,p}_j=\frac{\p u_i^{(p)}}{\p t^{\al,0}}\frac{\p v^\al}{\p u_j},\quad
i, j=1,\dots, n,\ p\ge 0.
\end{equation}
Then the
following recursion relation holds true
\begin{equation}\label{du-zh-3}
U^{i,p}_j= \p_xU^{i,p-1}_j-\sum_k \Gamma_{kj}^s u_{k,x}U^{i,p-1}_s,\quad
i, j=1,\dots,n,\ p\ge 1. 
\end{equation}
Using this recursion relation one can represent $U^{i,p}_j$ in terms of jets $u_i^{(m)}$ with $ m\ge 1$,
the rotation coefficients $\gamma_{ij}$ and the Lam\'e coefficients $h_i$, starting from $U^{i,0}_j=\delta^i_j u_{j,x}$. Such expressions will be useful in dealing with differential operators of the form
\[
\frac{\p v^\al}{\p u_i} \frac{\p}{\p t^{\al,0} }= \sum_{p\geq 0} U^{j,p}_i \frac{\p}{\p u_j^{(p)}}.
\]

The topological solution $v({\bf t})=\left(v^1({\bf t}),\dots, v^n({\bf t})\right)$ of the principal hierarchy
is determined from the system of $n$ equations
\[\sum \tilde{t}^{\al,p} \nabla \theta_{\al,p}=0,\quad \tilde{t}^{\al,p}=t^{\al,p}-\delta^\al_1 \delta^p_1.\]    
By using the topological solution $v({\bf t})$ one can define the genus 
zero free energy $\F_0=\F_0({\bf t})$ of the Frobenius manifold \cite{D1} satisfying
the equations
\begin{equation}\label{fzero-zh}
\frac{\p^3\F_0({\bf t})}{\p t^{\al,0}\p t^{\beta,0} \p t^{\gamma,0}}
=c_{\al\beta}^\xi(v({\bf t})) M_{\xi\gamma}, \quad \alpha,\beta,\gamma=1,\dots,n,
\end{equation}
where
\[M_{\xi\gamma}=c_{\xi\gamma\rho}(v({\bf t})) v^\rho_x.\]
\begin{rmk}
By taking $\alpha=1$ in \eqref{fzero-zh}, we see that the matrix $M_{\beta\gamma}$
coincides with the one appeared in the definition of the sixteen diagrams of Theorem \ref{main-thm},
so we use the same notation.
\end{rmk}
Observe the following useful formula for the entries of the inverse matrix
\begin{equation}\label{minv-zh}
\left( M^{-1}\right)^{\al\beta} = \sum_{i=1}^n \frac1{h_i^2 u_{i,x}} \frac{\p v^\al}{\p u_i} \frac{\p v^\beta}{\p u_i}.
\end{equation}
We also need to use the genus one free energy $\F_1({\bf t})$ which is defined for a semisimple Frobenius manifold by the following expression
\begin{equation}\label{def-f1-zh}
\F_1({\bf t})=\left.F_1(u,u_x)\right|_{v^\al=v^\al({\bf t})}\
\textrm{with}\ 
F_1(u,u_x)=\frac1{24} \sum_{i=1}^n \log{u_{i,x}}+G(u),
\end{equation}
where the function $G$ is called the G-function of the Frobenius manifold. It is given by a quadrature due to the following  
equations \cite{DZ-cmp}
\begin{equation}\label{dg-zh}
\frac{\p G(u)}{\p u_i}=\frac12 \sum_{j\ne i} (u_i-u_j) \gamma_{ij}^2-\frac1{24} \sum_{k\ne i} \gamma_{ik} \left(\frac{h_i}{h_k}-\frac{h_k}{h_i}\right).
\end{equation}

In order to write down the correlation functions in terms of canonical coordinates,
let us introduce the following notation
\begin{align*}
&C_{i_1,i_2,\dots,i_m}=\frac{\p^m\F_0({\bf t})}{\p t^{\al_1,0}\p t^{\al_2,0}\dots\p t^{\al_m,0}}\frac{\p v^{\al_1}}{\p u_{i_1}}\frac{\p v^{\al_2}}{\p u_{i_2}}\cdots \frac{\p v^{\al_m}}{\p u_{i_m}},\\
&D_{i_1,i_2,\dots,i_m}=\frac{\p^m\F_1({\bf t})}{\p t^{\al_1,0}\p t^{\al_2,0}\dots\p t^{\al_m,0}}\frac{\p v^{\al_1}}{\p u_{i_1}}\frac{\p v^{\al_2}}{\p u_{i_2}}\cdots \frac{\p v^{\al_m}}{\p u_{i_m}}
\end{align*}
for the indices $1\le i_1,\dots, i_m\le n$.
Then we have
\begin{equation}\label{cd-zh-1}
C_{i_1,i_2,i_3}=\left\{\begin{array}{ll} h_i^2\, u_{i_1,x},\quad &\textrm{if}\ i_1=i_2=i_3\\
0,\quad &\textrm{other cases}\end{array}\right.;\quad
D_{i}=\sum_{p=0}^1 U^{j,p}_i\,\frac{\p F_1(u, u_x)}{\p u^{(p)}_j}.
\end{equation}
By using the relation \eqref{chr-zh} we obtain the following recursive formula
\begin{align}
&X_{i_1,i_2,\dots,i_{m+1}}=\sum_{k=1}^n \sum_{p=0}^{m-2} \frac{\p X_{i_1,\dots,i_m}   }{\p u_k^{(p)}}\,
X^{k,p}_{i_{m+1}}\nn \\
&\qquad -\sum_{k=1}^m X_{i_1,\dots,i_{k-1},s,i_{k+1},\dots, i_m}
\Gamma^s_{i_k i_{m+1}} u_{i_{m+1},x}\label{recur-zh}
\end{align}                                 
which is valid for $X=C$ and $X=D$.

\noindent{\em{Proof of Theorem \ref{main-thm}.}} \ 
Since the genus two free energy $F_2$ given in \cite{DZ} is represented as a rational function of the canonical coordinates $u_i$, their $x$-derivatives $u^{(p)}_i=\p^p_x u_i$, the rotation coefficients $\gamma_{ij}$ and the Lam\'e coefficients $h_i$, 
in order to prove the theorem we need to represent the functions $Q_1,\dots, Q_{16}$ that are associated to the 16 dual graphs as rational functions of the  above mentioned variables. 
In fact, for the functions $Q_1$ and $Q_{16}$  defined in \eqref{q-1}, \eqref{q-16} we have
\begin{align*}
Q_1&=\frac{\p^6 \F_0}{\p t^{1,0}\,\p t^{1,0}\,\p t^{\alpha,0}\,\p t^{{\alpha'},0}\,\p t^{\beta,0}\,\p t^{{\beta'},0}}
 \sum_{j_1, j_2=1}^n \frac1{h_{j_1}^2 u_{j_1,x}} \frac{\p v^\al}{\p u_{j_1}} \frac{\p v^{\al'}}{\p u_{j_1}}
 \frac1{h_{j_2}^2 u_{j_2,x}} \frac{\p v^\beta}{\p u_{j_2}} \frac{\p v^{\beta'}}{\p u_{j_2}}\\
&=\sum_{i_1, i_2, j_1, j_2=1}^n \frac{\p v^{\alpha_1}}{\p u_{i_1}} 
\frac{\p v^{\alpha_2}}{\p u_{i_2}} \frac{\p^6 \F_0}{\p t^{\alpha_1,0}\,\p t^{\alpha_2,0}\,\p t^{\alpha,0}\,\p t^{\alpha',0}\,\p t^{\beta,0}\,\p t^{\beta',0}}\\
&\hskip 2cm \times 
\frac1{h_{j_1}^2 u_{j_1,x}} \frac{\p v^\al}{\p u_{j_1}} \frac{\p v^{\al'}}{\p u_{j_1}}
 \frac1{h_{j_2}^2 u_{j_2,x}} \frac{\p v^\beta}{\p u_{j_2}} \frac{\p v^{\beta'}}{\p u_{j_2}}\\
&=\sum_{i_1,i_2,j_1,j_2=1}^n \frac{C_{i_1, i_2, j_1, j_1, j_2, j_2}}{h_{j_1}^2 h_{j_2}^2 u_{j_1,x} u_{j_2,x}}\, ,
\end{align*}
and 
\begin{align*}
Q_{16}&=\frac{\p^2\F_1}{\p t^{1,0}\p t^{\al,0}} \sum_{i=1}^n \frac{1}{h_i^2 u_{i,x}} \frac{\p v^\al}{\p u_i}\frac{\p v^{\al'}}{\p u_i}\frac{\p\F_1}{\p  t^{{\al'},0}}\\
&=\sum_{i,j=1}^n \frac{\p^2\F_1}{\p t^{\beta,0}\p t^{\al,0}} \frac{\p v^\beta}{\p u_j} \frac{1}{h_i^2 u_{i,x}} \frac{\p v^\al}{\p u_i}\frac{\p v^{\al'}}{\p u_i}\frac{\p\F_1}{\p  t^{{\al'},0}}\\
&=\sum_{i,j=1}^n  \frac{D_i D_{i,j}}{h_i^2 u_{i,x}}.
\end{align*}
Here we used the identity
\[\sum_{i=1}^n \frac{\p v^\alpha}{\p u_i}=\frac{\p v^\alpha}{\p v^1}=\delta^\alpha_1.\] 
since the unit vector field $e$ of  the Frobenius manifold
equals $\frac{\p}{\p v^1}=\sum_{i=1}^n \frac{\p}{\p u_i}$.

From the formulae \eqref{dgamma-zh}--\eqref{recur-zh} it follows that the functions $C_{i_1, i_2, j_1, j_1, j_2, j_2}$, $D_{i}, D_{i,j}$ can also be represented 
as rational functions of the canonical coordinates $u_i$, their $x$-derivatives $u^{(p)}_i=\p^p_x u_i$, the rotation coefficients $\gamma_{ij}$ and the Lam\'e coefficients $h_i$.  In a similar way we can similar expressions for other functions $Q_2$, \dots $Q_{15}$. Now by subtracting the linear combination of the 16 functions $Q_1,\dots, Q_{16}$ that 
appear at the r.h.s of \eqref{mthm-zh} from the one 
given by the l.h.s. of \eqref{mthm-zh}, we get the needed expression for $G^{(2)}(u,u_x, u_{xx})$ by a tedious but straightforward computation. The theorem is proved.   

\subsection{Proof of Theorem \ref{thm-relation}} \label{sec-22}

In this section, we reduce the identity \eqref{relation} to an easier one \eqref{double}.

\begin{lem}\label{partial-x}
Let $\Gamma$ be a dual graph, and $x=t^{1,0}$. Then
\begin{equation}\label{ga-1}
\p_x \Gamma=\sum_{v:\,\mbox{\scriptsize vertex of } \Gamma} \Gamma_v-\sum_{e:\,\mbox{\scriptsize edge of } \Gamma} \Gamma_e,
\end{equation}
where $\Gamma_v$ is the dual graph obtained from $\Gamma$ by adding a new leg on the vertex $v$, and $\Gamma_e$ is the dual graph obtained from $\Gamma$ by adding a new vertex of genus zero with two legs on the edge $e$.
\end{lem}
\begin{prf}
The dual graph $\Gamma$ corresponds to the product of several multi-point correlation functions and the inverse of
the matrix $M$. According to the Leibniz rule, when the operator $\p_x$ acts on multi-point correlation functions, we
obtain terms that appear in the first summation of the r.h.s. of \eqref{ga-1}, and when it acts on the inverse of $M$,
we obtain terms that appear in the second summation. The lemma is proved.
\end{prf}

Let us introduce the following dual graphs:
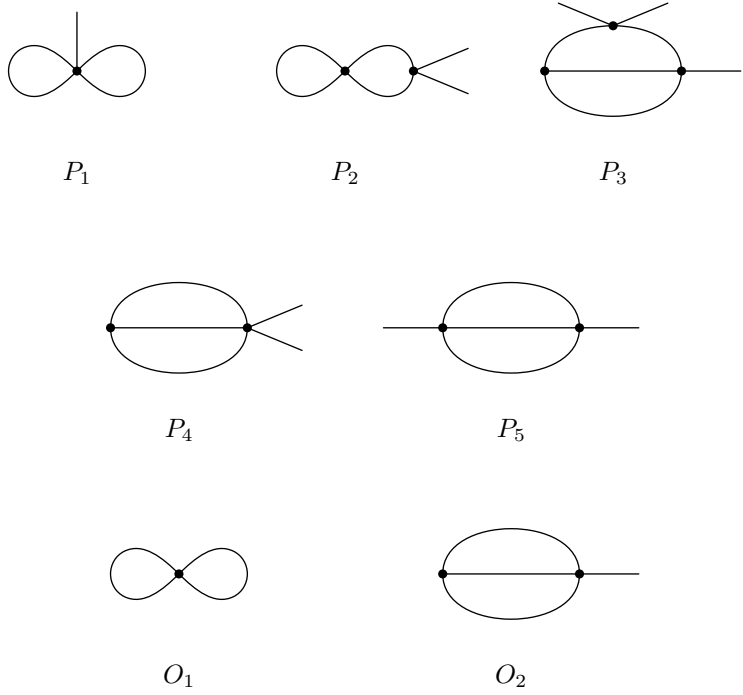
\[
\begin{array}{ccc}
\includegraphics{Q17} & \includegraphics{Q18} & \includegraphics{Q19}\\
P_1 & P_2 & P_3
\end{array}
\]
\[
\begin{array}{ccc}
\includegraphics{Q20} & \qquad & \includegraphics{Q21}\\
P_4 & & P_5
\end{array}
\]
\[
\begin{array}{ccc}
\includegraphics{Q22} & \qquad & \includegraphics{Q23}\\
O_1 & & O_2
\end{array}
\]
The we have the following lemma.
\begin{lem}
The following identities hold true
\begin{align*}
\p_x P_1=&Q_1-2\,Q_3,\\
\p_x P_2=&Q_3+Q_5-Q_7-2\,Q_9,\\
\p_x P_3=&Q_4+Q_8+Q_{10}-2\,Q_{11}-2\,Q_{12},\\
\p_x P_4=&Q_6+Q_2-3\,Q_{10},\\
\p_x P_5=&2\,Q_2-3\,Q_4,\\
\p_x O_1=&P_1-2\,P_2,\\
\p_x O_2=&P_4+P_5-3\,P_3,
\end{align*}
hence
\begin{align}
&(Q_1-Q_6)+2(Q_7-Q_5)+3(Q_8- Q_2)\nn\\
&\qquad\qquad+4(Q_9- Q_3)+6 (Q_4+Q_{10}-Q_{11}- Q_{12}) \nn\\
=&\p_x^2\left(O_1-O_2\right). \label{relation-2}
\end{align}
\end{lem}
\begin{prf}
They are easy consequences of Lemma \ref{partial-x}.
\end{prf}

\begin{lem}
For any semisimple Frobenius manifold, the following identity holds true
\begin{equation}\label{double}
O_1-O_2=\sum_{1\le i < j\le n}\gamma_{ij}\frac{(h_i^2+h_j^2)^2}{h_i^3\,h_j^3}.
\end{equation}
\end{lem}
\begin{prf}
The functions $O_1$, $O_2$ have the following expression.
\[O_1=\sum_{1\le j_1,j_2\le n} \frac{C_{ j_1, j_1, j_2, j_2}}{h_{j_1}^2 h_{j_2}^2 u_{j_1,x} u_{j_2,x}}\,,\quad O_2=\sum_{1\le j_1,j_2, j_3\le n}  \frac{C_{ i_1, j_1, j_2, j_3} C_{j_1, j_2, j_3}}{h_{j_1}^2 h_{j_2}^2 h_{j_3}^2 u_{j_1,x} u_{j_2,x} u_{j_3,x}}.\]
By using the formulae \eqref{cd-zh-1}, \eqref{recur-zh} one can obtain that
\begin{align*}
&O_1=\sum_{1\le i<j\le n} \gamma_{ij}\frac{(h_i^2\, u_{j,x}+h_j^2\, u_{i,x})^2-(h_i^4+h_j^4) (u_{i,x}-u_{j,x})^2}{h_i^3 h_j^3 u_{i,x} u_{j,x}}+\sum_{i=1}^n \frac{u_{i,xx}}
{h_i^2 u_{i,x}^2}\,,\\
&O_2=\sum_{1\le i<j\le n} \gamma_{ij}\frac{(h_i^4\, u_{i,x}-h_j^4\, u_{j,x})(u_{j,x}-u_{i,x})}
{h_i^3 h_j^3 u_{i,x} u_{j,x}}+\sum_{i=1}^n \frac{u_{i,xx}}
{h_i^2 u_{i,x}^2}\,.
\end{align*}
Then one can easily see that the difference $O_1-O_2$ equals the r.h.s. of \eqref{double}.
The lemma is proved.
\end{prf}

To prove Theorem \ref{thm-relation} one only need to prove the following lemma.
\begin{lem}\label{easier-lem}
For a Frobenius manifold associated to ADE singularities, or $\PP^1$-orbifolds of AD type,
the difference $O_1-O_2$ is always a constant.
\end{lem}
We will give the proof of the above lemma case by case in Section \ref{sec-3}.

\subsection{Rotation coefficients for simple singularities} \label{sec-23}
 
Let $f$ be a polynomial on $\C^m$ which has an isolated critical point at $0 \in \C^m$ of $ADE$ type. Let $n$ be the Milnor number of $f$.
The coordinates in $\C^m$ are $z=(z^1, \dots, z^m)$. We denote $\p_\alpha$ or $\p_{z^\alpha}$ the partial derivatives $\frac{\p}{\p z^\alpha}$.

Let $F: \C^m \times B \to \C,\ (z, t) \mapsto F(z,t)$ be a miniversal unfolding of $f$ (avoid confusions with the potential of the Frobenius manifold!) where $B$ is an open ball in $\C^n$.
Let $C\subset B$ be the caustic. For a given point $t$ in the complement $B\setminus C$ the function $F(z,t)$ has $n$ Morse critical points
$z^{(i)}(t)=(z^{(i),1},\dots, z^{(i),m}) \ (i=1,\dots,n)$,
\[\p_\alpha F(z,t)|_{z=z^{(i)}(t)}=0,\ \alpha=1, \dots, m.\]
Define the canonical coordinates $u_i$ on $B\setminus C$ as the critical values
\begin{equation}\label{can}
u_i(t)=F(z^{(i)}(t),t),\quad i=1,\dots, n.
\end{equation}
We will often use short notations  $\p_i$ or $\p_{u_i}$ for the partial derivatives $\frac{\p}{\p u_i}$.

There is a semisimple Frobenius manifold structure on the base space $B\setminus C$.
The flat metric $\la\ ,\ \ra$ is defined by
\begin{equation}\label{residue-pairing}
\la \p', \p''\ra_t = -\res_{z=\infty}\frac{(\p' F(z,t))(\p'' F(z,t))\ dz^1\wedge\cdots\wedge dz^m}{\p_{z^1} F
\cdots \p_{z^m} F}
\end{equation}
for any $\p',\ \p''\in T_tB$.
We denote $h_{\alpha\beta}(z,t)=\p_\alpha\p_\beta F(z,t)$, $H(z,t)=\det(h_{\alpha\beta}(z,t))$.
Let $(h^{\alpha\beta})$ be the inverse matrix of $(h_{\alpha\beta})$.
Then from the residue theorem it follows that

\begin{equation}
\la \p', \p''\ra_t=\sum_{k=1}^n \left.\frac{(\p' F(z,t))(\p'' F(z,t))}{H(z,t)}\right|_{z=z^{(k)}(t)}. \label{eq-1}
\end{equation}

Denote
\begin{equation}\label{eta}
\eta^{ii}(t)=H(z^{(i)}(t),t), \quad \eta_{ii}(t)=\left( H(z^{(i)}(t),t)\right)^{-1}.
\end{equation}
Then by using \eqref{eq-1} and the identity 
\begin{equation}
\p_i F(z,t)|_{z=z^{(k)}(t)}=\delta_{ik} \label{eq-2}
\end{equation}
we obtain
\begin{equation}\label{eq-3}
\la \p_i, \p_j\ra_t=\sum_{k=1}^n \left.\frac{(\p_i F(z,t))(\p_j F(z,t))}{H(z,t)}\right|_{z=z^{(k)}(t)}
=\sum_{k=1}^n \frac{\delta_{ik}\delta_{jk}}{\eta^{kk}(t)}=\delta_{ij}\,\eta_{ii}(t).
\end{equation}
From the definition of the critical points $z^{(k)}(t)$ it follows that
\begin{align}
&\p_i\p_\alpha F(z,t)|_{z=z^{(k)}(t)}=-h_{\alpha\beta}(z^{(k)}(t),t)\,\p_i z^{(k),\beta}(t). \label{eq-41}\\
&\p_i z^{(k),\beta}(t)=-h^{\alpha\beta}(z^{(k)}(t),t)\,\p_i\p_\alpha F(z,t)|_{z=z^{(k)}(t)}. \label{eq-42}
\end{align}
By using these equations and the identity
\[\p_x \det A(x)=\det A(x) \tr\left(A^{-1}(x) \p_x A(x) \right)\]
for any nondegenerate matrix function $A(x)$, we obtain
\begin{align}
\frac{\p_i\eta^{kk}}{\eta^{kk}}=&\left(h^{\alpha\beta}(z,t)\p_ih_{\alpha\beta}(z,t)-h^{\alpha\beta}(z,t)
\p_\gamma h_{\alpha\beta}(z,t)\,h^{\gamma\sigma}(z,t)\,\p_i\p_\sigma F(z,t)\right)|_{z=z^{(k)}(t)}\nn\\
=&\left(h^{\alpha\beta}(z,t)\p_ih_{\alpha\beta}(z,t)+\p_\alpha h^{\alpha\sigma}\,\p_i\p_\sigma F(z,t)\right)|_{z=z^{(k)}(t)}\nn\\
=&\p_\alpha\left(h^{\alpha\beta}(z,t)\,\p_i\p_\beta F(z,t)\right)|_{z=z^{(k)}(t)}.\label{eq-5}
\end{align}

As above we denote $h_i=\sqrt{\eta_{ii}}$ the Lam\'e coefficients and
$
\gamma_{ki}=\frac{\p_i\,h_k}{h_i}
$
the rotation coefficients of the metric $\sum_{i=1}^n \eta_{ii}(du_i)^2$.
Of the Christoffel symbols of the metric we will often use the coefficients $\Gamma_{ki}^k$ with $k\neq i$, so we introduce a notation for these coefficients
\begin{equation}\label{gamma-1}
\Gamma_{ki}:=\Gamma^k_{ki}=\frac{\p_i\eta_{kk}}{2\,\eta_{kk}}=-\frac12\p_\alpha\left(h^{\alpha\beta}(z,t)\,\p_i\p_\beta F(z,t)\right)|_{z=z^{(k)}}.
\end{equation}
Then
\begin{equation}\label{gamma-2}
\gamma_{ki}=\frac{h_k}{h_i}\,\Gamma_{ki}.
\end{equation}

\begin{rmk}
The equations \eqref{dg-zh} satisfied by the G-function of the Frobenius manifold can be rewritten as
\begin{equation}
\p_iG(u)=\frac12\sum_{k \ne i}(u_i-u_k)\Gamma_{ki}\Gamma_{ik}-\frac1{24}\sum_{k \ne i} (\Gamma_{ki}-\Gamma_{ik}). \label{eq-6}
\end{equation}
The explicit expressions of $\Gamma_{ki}$ given in Sec.\,\ref{sec-3} for the Frobenius manifolds associated to ADE singularities can be used to re-derive the known explicit formulae $G=0$ \cite{givental, stra} for the G-functions of this class of Frobenius manifolds. We can also obtain the explicit formulae \eqref{G-func-A},
\eqref{G-func-D} for the G-functions of the Frobenius manifolds 
defined on the orbit spaces of the extended affine Weyl groups of AD type. I.\,Strachan proved the formula \eqref{G-func-A} (see below) and conjectured the formula \eqref{G-func-D} in \cite{stra}.  
\end{rmk}

The equations \eqref{gamma-1} and \eqref{gamma-2} give us a formula for computing the rotation coefficients of the
Frobenius manifold. However, one also needs to compute the derivatives of $F(z,t)$ w.r.t. the canonical
coordinates. To this end, starting from this point we assume that the miniversal deformation $F(z,t)$ is given by
\[F(z,t)=f(z)+\sum_{j=1}^n t^j\,\phi_j(z),\]
where $\phi_1(z), \dots, \phi_n(z)$ is a basis of the Milnor ring. Define $W:(\C^m)^n\to\C$
\[W(z_1, \dots, z_n)=\det(\phi_j(z_i)).\]

\begin{lem}\label{lem-7}
\begin{equation}
\p_i F(z,t)=\frac{W(z^{(1)}, \dots, z^{(i-1)}, z, z^{(i+1)}, \dots, z^{(n)})}{W(z^{(1)}, \dots, z^{(n)})}. \label{eq-7}
\end{equation}
\end{lem}
\begin{prf}
From  \eqref{eq-2} it follows that
\[\sum_{j=1}^n \frac{\p t^j}{\p u_i}\,\phi_j(z^{(k)}(t))=\delta_{ik}.\]
So we have
\[\frac{\p u_i}{\p t^j}=\phi_j(z^{(i)}(t)).\]
Next,  let us consider the following system of linear equations for partial derivatives $\p_i F(z,t)=\frac{\p F(z,t)}{\p u_i}$
\[\phi_j(z)=\frac{\p F(z,t)}{\p t^j}=\p_i F(z,t)\,\frac{\p u_i}{\p t^j}=\phi_j(z^{(i)}) \p_i F(z,t),\ j=1, \dots, n.\]
The statement of the lemma now follows by using Cramer's rule.
\end{prf}

\subsection{Rotation coefficients for $\PP^1$-orbifolds} \label{sec-24}

Let $p,q,r$ are positive integers satisfying
\[\frac1p+\frac1q+\frac1r>1.\]
It is shown in \cite{rossi} that the quantum cohomology of the $\PP^1$-orbifold $\PP^1_{p,q,r}$ is isomorphic
to the Frobenius structure on the space of  tri-polynomials of type $(p,q,r)$.

We take $m=3$, $n=p+q+r-1$. A tri-polynomial is a function $F: \C^m \times B \to \C,\ (z, t) \mapsto F(z,t)$, 
\begin{align}
&F(z,t)=-z^1z^2z^3+P_1(z_1)+P_2(z_2)+P_3(z_3), \\
&\qquad P_1(z_1)=\sum_{i=1}^{p-1}t_i z_1^i+z_1^p,\\
&\qquad P_2(z_2)=\sum_{i=1}^{q-1}t_{p-1+i} z_2^i+z_2^p,\\
&\qquad P_3(z_3)=\sum_{i=0}^rt_{p+q-1+i} z_3^i,
\end{align}
where $B$ is an open set in $\C^{n-1}\times\C^*$ defined by the condition $t^n\ne0$. 
Let $C\subset B$ be the caustic. Like in the previous section the critical values
\begin{equation}\label{can}
u_i(t)=F(z^{(i)}(t),t),\quad i=1,\dots, n.
\end{equation} 
define the canonical coordinates $u^i$
on $B\setminus C$.

The flat metric of the Frobenius structure on the space of tri-polynomial is also defined by \eqref{residue-pairing}.
One can easily see that all lemmas from the previous section hold true also for tri-polynomials.

\section{Examples} \label{sec-3}

\subsection{The $A_n$ singularities} \label{sec-31}

In this case, $m=1$, $f(z)=z^{n+1}$, $\phi_j=z^{n-j}$.

\begin{lem}\label{lem-8}
\begin{equation}
\p_i F(z,t)=\frac1{z-z^{(i)}}\frac{F'(z,t)}{F''(z^{(i)},t)}. \label{eq-8}
\end{equation}
\end{lem}
\begin{prf}
By using the identities
\begin{align*}
F'(z,t)=&(n+1)\prod_{k=1}^n (z-z^{(k)}(t)),\\
F''(z^{(i)}(t),t)=&(n+1)\prod_{k\ne i} (z^{(i)}(t)-z^{(k)}(t))
\end{align*}
and Lemma \ref{lem-7}, the lemma can be easily proved.
\end{prf}

\begin{lem}\label{lem-9}
\begin{equation}
\Gamma_{ki}(t)=\frac1{(z^{(k)}(t)-z^{(i)}(t))^2 F''(z^{(i)}(t),t)}. \label{eq-9}
\end{equation}
\end{lem}
\begin{prf}
It follows from \eqref{eq-5} and Lemma \ref{lem-8}.
\end{prf}

\begin{rmk}
By applying the residues theorem to the following two meromorphic functions
\[m(z)=\frac{F(z)-F(z^{(i)})}{F'(z)(z-z^{(i)})^4},\ \tilde{m}(z)=\frac{F''(z)-F''(z^{(i)})}{F'(z)(z-z^{(i)})^2},\]
one can easily prove that the G-functions of $A_n$ singularities vanish.
\end{rmk}

Now let us use the formula \eqref{eq-9} to verify the validity of Conjecture \ref{main-conj1} for $A_n$ singularities.
We use the critical points $z^{(1)},\dots, z^{(n)}$ and an additional parameter $z^{(0)}$ to represent $F(z,t)=z^{n+1}+t^1 z^{n-1}+\dots+t^n$ as
\begin{equation}
F(z,t)=\lambda(z)=\int_0^z (n+1)\prod_{k=1}^{n}(\xi-z^{(k)}) d\xi+z^{(0)}
\end{equation}
Note that $z^{(1)},\dots z^{(n)}$ are not independent as they satisfy
\begin{equation}\label{sumzero}
z^{(n)}=-\sum_{k=1}^{n-1} z^{(k)}.
\end{equation}
We have
\begin{equation}\label{ga-A}
u_i=\lambda(z^{(i)}),\quad h_i=\psi_{i,1}=\frac1{\sqrt{\lambda''(z^{(i)})}},\quad 
\gamma_{ij}=\frac{h_i\,h_j}{(z^{(i)}-z^{(j)})^2}.
\end{equation}
By substituting these expressions into the formula \eqref{g2-zh} for $G^{(2)}(u,u_x,u_{xx})$, we obtain a rational function of
$z^{(0)},\dots z^{(n-1)}$. For $n\le 8$  one can check with the help of a suitable symbolic computations software that this rational
functions vanishes, so the Conjecture \ref{main-conj1} holds true for such cases.

\vskip 1em

\begin{prfn}{Lemma \ref{easier-lem} for $A_n$ singularities.}
Let us denote $z^{(i)}$ by $z_i$ for $i=1,\dots, n$. By using the formulae given in \eqref{ga-A} we obtain
\begin{align*}
\sum_{1\le i < j\le n}\gamma_{ij}\frac{(h_i^2+h_j^2)^2}{h_i^3\,h_j^3}=&
\sum_{1\le i < j\le n}\frac{1}{(z_i-z_j)^2}\left(\frac{\lambda''(z_i)}{\lambda''(z_j)}+\frac{\lambda''(z_j)}{\lambda''(z_i)}+2\right) \\
=&\sum_{i=1}^n \sum_{j\ne i}\frac{\lambda''(z_i)+\lambda''(z_j)}{(z_i-z_j)^2\,\lambda''(z_j)}.
\end{align*}
For fixed $i$ one has
\begin{align*}
&\sum_{j\ne i}\frac{\lambda''(z_i)+\lambda''(z_j)}{(z_i-z_j)^2\,\lambda''(z_j)}=\sum_{j\ne i}\res_{z=z_j}\frac{\lambda''(z)+\lambda''(z_i)}{(z-z_i)^2\,\lambda'(z)}\\
=&-\res_{z=z_i}\frac{\lambda''(z)+\lambda''(z_i)}{(z-z_i)^2\,\lambda'(z)}=-\frac16\frac{\lambda^{(4)}(z_i)}{\lambda''(z_i)}.
\end{align*}
So
\begin{align*}
&\sum_{i=1}^n \sum_{j\ne i}\frac{\lambda''(z_i)+\lambda''(z_j)}{(z_i-z_j)^2\,\lambda''(z_j)}=-\frac16\,\sum_{i=1}^n\frac{\lambda^{(4)}(z_i)}{\lambda''(z_i)}\\
=&-\frac16\,\sum_{i=1}^n\res_{z=z_i}\frac{\lambda^{(4)}(z)}{\lambda'(z)}=\frac16\,\res_{z=\infty}\frac{\lambda^{(4)}(z)}{\lambda'(z)}=0\,.
\end{align*}
The lemma is proved.
\end{prfn}

\subsection{The $D_n$ singularities} \label{sec-32}

In this case, $m=2$. Denote $x=z^1, y=z^2$ and $f(z)=x^{n-1}+x\,y^2$. A basis in the Milnor ring is given by 
\[\phi_j=x^{n-j-1}\ (j=1, \dots, n-1),\ \phi_n=y.\]
The critical points are determined from the following equations
\begin{align*}
F_x=&(n-1)x^{n-2}+\cdots+t^{n-2}+y^2=0,\\
F_y=&2\,x\,y+t^n=0,
\end{align*}
or, equivalently,
\[y=-\frac{t^n}{2\,x},\ (n-1)x^{n-2}+\cdots+t^{n-2}+\frac{(t^n)^2}{4\,x^2}=0.\]
We introduce a function
\[\lambda(x,t)=x^{n-1}+\sum_{j=1}^{n-1}t^j\phi_j-\frac{(t^n)^2}{4\,x},\]
then the critical points and the critical values of $F(z,t)$ are given by the ones of $\lambda(x,t)$. We denote $z^{(i)}=(x_i,y_i)$,

\begin{lem} \label{lem-11}
\begin{equation}
\p_iF(z,t)=\frac1{x-x_i}\frac{x}{x_i}\frac{\lambda'(x)}{\lambda''(x_i)}+\frac{t^n (2\,x\,y+t^n)}{4\,x\,x_i^2\lambda''(x_i)}. \label{eq-11}
\end{equation}
\end{lem}
\begin{prf}
We need to compute the denominator and the numerator of the right hand side of \eqref{eq-7}.

Since $y_i=-\frac{t^n}{2\,x_i}$, the denominator can be converted to a Vandermonde determinant
\[W(z^{(1)}, \dots, z^{(n)})=-\frac{t^n}{2\,x_1 \cdots x_n}\prod_{1 \le k < l \le n}(x_k-x_l).\]

To compute the numerator, we rewrite $y$ as
\[y=\left(-\frac{t^n}{2\,x}\right)+\left(y+\frac{t^n}{2\,x}\right),\]
then split the determinant into two parts,
\begin{align*}
&W(z^{(1)}, \dots, z^{(i-1)}, z, z^{(i+1)}, \dots, z^{(n)})\\
=&\left|\begin{array}{cccccc}
x_1^{n-2} & x_1^{n-3} & \cdots & x_1 & 1 & -\frac{t^n}{2\,x_1} \\
\cdots & \cdots & \cdots & \cdots & \cdots & \cdots \\
x^{n-2} & x^{n-3} & \cdots & x & 1 & -\frac{t^n}{2\,x} \\
\cdots & \cdots & \cdots & \cdots & \cdots & \cdots \\
x_n^{n-2} & x_n^{n-3} & \cdots & x_n & 1 & -\frac{t^n}{2\,x_n}
\end{array}\right|+\left|\begin{array}{cccccc}
x_1^{n-2} & x_1^{n-3} & \cdots & x_1 & 1 & -\frac{t^n}{2\,x_1} \\
\cdots & \cdots & \cdots & \cdots & \cdots & \cdots \\
0 & 0 & \cdots & 0 & 0 & y+\frac{t^n}{2\,x} \\
\cdots & \cdots & \cdots & \cdots & \cdots & \cdots \\
x_n^{n-2} & x_n^{n-3} & \cdots & x_n & 1 & -\frac{t^n}{2\,x_n}
\end{array}\right|.
\end{align*}
The first determinant is similar to the denominator, while the second one, by the Laplace expansion, is again a Vandermonde determinant, so we have
\begin{align*}
&W(z^{(1)}, \dots, z^{(i-1)}, z, z^{(i+1)}, \dots, z^{(n)})\\
=&-\frac{t^n}{2\,x_1 \cdots x_n}\frac{x_i}{x}\prod_{1 \le k < l \le n}(x_k-x_l)\ \prod_{k \ne i}\frac{x-x_k}{x_i-x_k}\\
&-(-1)^{n}\left(y+\frac{t^n}{2\,x}\right)\prod_{1 \le k < l \le n}(x_k-x_l)\ \prod_{k \ne i}\frac1{x_i-x_k}.
\end{align*}

By using Lemma \ref{lem-7}, we have
\[\p_iF(z,t)=\frac{x_i}{x}\prod_{k \ne i}\frac{x-x_k}{x_i-x_k}+(-1)^n\frac{2\,x_1 \cdots x_n}{t^n\ \prod_{k \ne i}(x_i-x_k)}
\left(y+\frac{t^n}{2\,x}\right).\]
With the help the following simple identities
\begin{align*}
&\frac{x^2\,\lambda'(x)}{n-1}=\prod_{k=1}^n(x-x_k),\ \frac{x_i^2\,\lambda''(x_i)}{n-1}=\prod_{k \ne i}^n(x_i-x_k),\\
&x_1 \cdots x_n=(-1)^n\frac{(t^n)^2}{4(n-1)},
\end{align*}
the proof of the lemma can be completed in a straightforward way.
\end{prf}

\begin{lem} \label{lem-12}
\begin{equation}
\Gamma_{ki}=\frac{x_k+x_i}{(x_k-x_i)^2\,2\,x_i\,\lambda''(x_i)}. \label{eq-12}
\end{equation}
\end{lem}
\begin{prf}
It follows from eq. \eqref{eq-5} and Lemma \ref{lem-11}.
\end{prf}

\begin{rmk}
By computing residues of the meromorphic functions
\[m(x)=\frac{(\lambda(x)-\lambda(x_i))(x+x_i)^2}{(x-x_i)^4\,x\,\lambda'(x)},\ 
\tilde{m}(x)=\frac{(x\,\lambda''(x)-x_i\,\lambda''(x_i))(x+x_i)}{(x-x_i)^2\,x\,\lambda'(x)},\]
one can easily prove that the G-functions of $D_n$ singularities vanish.
\end{rmk}

To verify Conjecture \ref{main-conj1} for $D_n$ singularities, let us represent $\lambda(x)=\lambda(x,t)$ in terms of $x_1,\dots, x_{n-1}$ and $x_0$
in the form
\begin{equation}
\lambda(x)=\int_0^x (n-1) \xi^{-2} \prod_{k=1}^n (\xi-x_i) d\xi+x_0
\end{equation}
Here $\frac1{x_n}=-\sum_{k=1}^{n-1} \frac1{x_k}$. Then we have
\begin{equation}\label{ga-D}
u_i=\lambda(x_{i}),\quad h_i=\psi_{i,1}=\frac1{\sqrt{2 x_i \lambda''(x_i)}},\quad 
\gamma_{ij}=\frac{(x_i+x_j) h_i h_j}{(x_i-x_j)^2}.
\end{equation}
By using these data, one can verify the Conjecture \ref{main-conj1} for the $D_n$ singularity case
for  $n\le 8$.

\vskip 1em

\begin{prfn}{Lemma \ref{easier-lem} for $D_n$ singularities}
From \eqref{ga-D} it follows the formula
\[
\sum_{1\le i < j\le n}\gamma_{ij}\frac{(h_i^2+h_j^2)^2}{h_i^3\,h_j^3}=
\sum_{i=1}^n \sum_{j\ne i}\frac{x_i+x_j}{(x_i-x_j)^2}\frac{x_i\,\lambda''(x_i)+x_j\,\lambda''(x_j)}{x_j\,\lambda''(x_j)}.
\]
Now For fixed $i$ we have
\begin{align*}
&\sum_{j\ne i}\frac{x_i+x_j}{(x_i-x_j)^2}\frac{x_i\,\lambda''(x_i)+x_j\,\lambda''(x_j)}{x_j\,\lambda''(x_j)}\\
=&\sum_{j\ne i}\res_{z=x_j}\frac{z+x_i}{(z-x_i)^2}\frac{z\,\lambda''(z)+x_i\,\lambda''(x_i)}{z\,\lambda'(z)}\\
=&-\left(\res_{z=0}+\res_{z=x_i}\right)\frac{z+x_i}{(z-x_i)^2}\frac{z\,\lambda''(z)+x_i\,\lambda''(x_i)}{z\,\lambda'(z)}\\
=&\frac2{x_i}-\left(\frac1{x_i}+\frac{\lambda'''(x_i)}{\lambda''(x_i)}+\frac{3\,x_i\,\lambda^{(4)}(x_i)}{\lambda''(x_i)}\right)\\
=&\frac1{x_i}-\frac{\lambda'''(x_i)}{\lambda''(x_i)}-\frac{3\,x_i\,\lambda^{(4)}(x_i)}{\lambda''(x_i)},
\end{align*}
so we obtain
\begin{align*}
&\sum_{i=1}^n \sum_{j\ne i}\frac{x_i+x_j}{(x_i-x_j)^2}\frac{x_i\,\lambda''(x_i)+x_j\,\lambda''(x_j)}{x_j\,\lambda''(x_j)}\\
=&\sum_{i=1}^n \left(\frac1{x_i}-\frac{\lambda'''(x_i)}{\lambda''(x_i)}-\frac{3\,x_i\,\lambda^{(4)}(x_i)}{\lambda''(x_i)}\right)\\
=&\sum_{i=1}^n \frac1{x_i}+\left(\res_{z=0}+\res_{z=\infty}\right)\left(\frac{\lambda'''(z)}{\lambda'(z)}+\frac{3\,z\,\lambda^{(4)}(z)}{\lambda'(z)}\right)\\
=&0+0+0+0+0=0\,.
\end{align*}
The lemma is proved.
\end{prfn}

\subsection{The $E_6$ and $E_8$ singularities} \label{sec-33}

In this case $m=2$ and
\begin{align*}
E_6 :\ & f(x,y)=x^3+y^4, \\
E_8 :\ & f(x,y)=x^3+y^5.
\end{align*}
Let $\nu=n/2$, then $f(x,y)=x^3+y^{\nu+1}$, and the miniversal deformation $F$ reads
\begin{equation}
F(z,t)=x^3+p(y)\,x +q(y),
\end{equation}
where
\[p(y)=\sum_{k=1}^\nu t_k\, y^{\nu-k},\ q(y)=y^{\nu+1}+\sum_{k=1}^\nu t_{\nu+k}\, y^{\nu-k}.\]
Here the indices of $t$'s are written as subscripts for convenience. 
The critical points are determined from the following equations
\begin{align*}
F_x=&3\,x^2+p(y)=0,\\
F_y=&p'(y)\,x+q'(y)=0.
\end{align*}
So we have $x=-q'(y)/p'(y)$, and
\[R(y):=R(F_x, F_y, x)=3\,q'(y)^2+p(y)p'(y)^2=0.\]
Here and below $R(f_1(u), f_2(u), u)$ stands for the resultant of polynomials $f_1$ and $f_2$ with respect to the variable $u$.
$R(y)$'s roots give us the $y$-components of all the critical points $z^{(k)}=(x_k, y_k)\ (k=1, \dots, n)$.
The corresponding $x$-components $x_k$'s can be obtained from \[
x_k=-\frac{q'(y_k)}{p'(y_k)}, \quad k=1, \dots, n.
\]

\begin{lem}\label{lem-14}
Let $\Delta=R(q'(y),p'(y),y)$. Then
\begin{equation}
W(z^{(1)}, \dots, z^{(n)})=(-1)^\nu\frac{(\nu+1)^{2\nu-2}}{\Delta}\prod_{1 \le k < l \le n}(y_k-y_l). \label{eq-14}
\end{equation}
\end{lem}
\begin{prf}
By definition
\[W(z^{(1)}, \dots, z^{(n)})=\left|
\begin{array}{ccccccccccccccc}
x_1\,y_1^{\nu-1} & & x_1\,y_1^{\nu-2} & & \cdots & & x_1 & & y_1^{\nu-1} & & y_1^{\nu-2} & & \cdots & & 1\\
\cdots & & \cdots & & \cdots & & \cdots & & \cdots & & \cdots & & \cdots & & \cdots\\
\cdots & & \cdots & & \cdots & & \cdots & & \cdots & & \cdots & & \cdots & & \cdots\\
x_{2\nu}\,y_{2\nu}^{\nu-1} & & x_{2\nu}\,y_{2\nu}^{\nu-2} & & \cdots & & x_{2\nu} & & y_{2\nu}^{\nu-1} & & y_{2
\nu}^{\nu-2}
& & \cdots & & 1
\end{array}\right|.\]
So we have
\begin{align*}
&(-1)^\nu\,\left(\prod_{k=1}^n p_k'\right)\,W(z^{(1)}, \dots, z^{(n)})\nn\\
=&\left|
\begin{array}{ccccccccccccccc}
q_1'\,y_1^{\nu-1} & & q_1'\,y_1^{\nu-2} & & \cdots & & q_1' & & p_1'\,y_1^{\nu-1} & & p_1'\,y_1^{m-2} & & \cdots & & p_1'\\
\cdots & & \cdots & & \cdots & & \cdots & & \cdots & & \cdots & & \cdots & & \cdots\\
\cdots & & \cdots & & \cdots & & \cdots & & \cdots & & \cdots & & \cdots & & \cdots\\
q_{2\nu}'\,y_{2\nu}^{\nu-1} & & q_{2\nu}'\,y_{2\nu}^{\nu-2} & & \cdots & & q_{2\nu}' & & p_{2\nu}'\,y_{2\nu}^{\nu-1} & &
p_{2\nu}'\,y_{2nu}^{\nu-2} & & \cdots & & p_{2\nu}'
\end{array}\right|, \\
=&|U\cdot V|=|U|\cdot |V|.
\end{align*}
Here $p'_i=p'(y_i)$ and $q'_i=q'(y_i)$, and the matrices $U$, $V$ read
\[U=\left(\begin{array}{ccccc}
y_1^{n-1} & y_1^{n-2} & \cdots & y_1 & 1 \\
\vdots & \vdots & & \vdots & \vdots \\
y_n^{n-1} & y_n^{n-2} & \cdots & y_n & 1
\end{array}\right),\]
\[
V=\left\{
\begin{array}{ll}
\left(\begin{array}{cccccc}
4 & 0 & 0 & 0 & 0 & 0 \\
0 & 4 & 0 & 0 & 0 & 0 \\
2\,t_4 & 0 & 4 & 2\,t_1 & 0 & 0 \\
t_5 & 2\,t_4 & 0 & t_2 & 2\,t_1 & 0 \\
0 & t_5 & 2\,t_4 & 0 & t_2 & 2\,t_1 \\
0 & 0 & t_5 & 0 & 0 & t_2
\end{array}\right), & \nu=3, \\ & \\
\left(\begin{array}{cccccccc}
5 & 0 & 0 & 0 & 0 & 0 & 0 & 0 \\
0 & 5 & 0 & 0 & 0 & 0 & 0 & 0 \\
3\,t_5 & 0 & 5 & 0 & 3\,t_1 & 0 & 0 & 0 \\
2\,t_6 & 3\,t_5 & 0 & 5 & 2\,t_2 & 3\,t_1 & 0 & 0 \\
t_7 & 2\,t_6 & 3\,t_5 & 0 & t_3 & 2\,t_2 & 3\,t_1 & 0 \\
0 & t_7 & 2\,t_6 & 3\,t_5 & 0 & t_3 & 2\,t_2 & 3\,t_1\\
0 & 0 & t_7 & 2\,t_6 & 0 & 0 & t_3 & 2\,t_2 \\
0 & 0 & 0 & t_7 & 0 & 0 & 0 & t_3
\end{array}\right), & \nu=4.
\end{array}
\right.\]
The matrix $U$ is just the Vandermonde matrix of $y_1, \dots, y_n$, so
\[|U|=\prod_{1 \le k < l \le n}(y_k-y_l).\]
Then by the determinant formula for the resultant $R(q'(y),p'(y),y)=\Delta$ one obtain.
\[|V|=(\nu+1)^2\,\Delta.\]
On the other hand, according to the properties of resultant, we have
\[\prod_{k=1}^n p_k'=\left(\frac{\Delta}{(\nu+1)^{\nu-2}}\right)^2.\]
So the lemma is proved.
\end{prf}

\begin{lem} \label{lem-15}
We have
\begin{equation}
\p_iF=\frac1{(y-y_i)\,R'(y_i)}\frac{p'(y_i)}{p'(y)}\Big(R(y)-3\,F_y(x,y)\,\Sigma\Big) \label{eq-15}
\end{equation}
where $\Sigma$ reads
\[
\Sigma=\left\{\begin{array}{ll}
F_y(x_i, y), & \nu=3, \\
F_y(x_i, y)+\frac{t_1^2}{5}(y-y_i)p'(y), & \nu=4.
\end{array}\right.\]
\end{lem}
\begin{prf}
According to Lemma \ref{lem-7}, $\p_i F=W_2/W_1$, where
\[W_1=W(z^{(1)}, \dots, z^{(n)}),\ W_2=W(z^{(1)}, \dots, z^{(i-1)}, z, z^{(i+1)}, \dots, z^{(n)}).\]
We now have to compute $W_2$.

First we rewrite $W_2$ in the following form
\[W_2=A(x-\tilde{x})+B,\]
where $\tilde{x}=-q'(y)/p'(y)$, and
\begin{align*}
A=&\left|
\begin{array}{ccccccccccccccc}
x_1\,y_1^{\nu-1} & & x_1\,y_1^{\nu-2} & & \cdots & & x_1 & & y_1^{\nu-1} & & y_1^{\nu-2} & & \cdots & & 1\\
\cdots & & \cdots & & \cdots & & \cdots & & \cdots & & \cdots & & \cdots & & \cdots\\
y^{\nu-1} & & y^{\nu-2} & & \cdots & &  1 & & 0 & & 0 & & \cdots & & 0\\
\cdots & & \cdots & & \cdots & & \cdots & & \cdots & & \cdots & & \cdots & & \cdots\\
x_{2\nu}\,y_{2\nu}^{\nu-1} & & x_{2\nu}\,y_{2\nu}^{\nu-2} & & \cdots & & x_{2\nu} & & y_{2\nu}^{\nu-1} & & y_{2
\nu}^{\nu-2}
& & \cdots & & 1
\end{array}\right|, \\ & \\
B=&\left|
\begin{array}{ccccccccccccccc}
x_1\,y_1^{\nu-1} & & x_1\,y_1^{\nu-2} & & \cdots & & x_1 & & y_1^{\nu-1} & & y_1^{\nu-2} & & \cdots & & 1\\
\cdots & & \cdots & & \cdots & & \cdots & & \cdots & & \cdots & & \cdots & & \cdots\\
\tilde{x}\,y^{\nu-1} & & \tilde{x}\,y^{\nu-2} & & \cdots & &  \tilde{x}\, & & y^{\nu-1} & & y^{\nu-2} & & \cdots & & 1\\
\cdots & & \cdots & & \cdots & & \cdots & & \cdots & & \cdots & & \cdots & & \cdots\\
x_{2\nu}\,y_{2\nu}^{\nu-1} & & x_{2\nu}\,y_{2\nu}^{\nu-2} & & \cdots & & x_{2\nu} & & y_{2\nu}^{\nu-1} & & y_{2
\nu}^{\nu-2}
& & \cdots & & 1
\end{array}\right|.
\end{align*}

The determinant $B$ is very similar to $W_1$, so one can obtain that
\[B=\frac1{(y-y_i)\,R'(y_i)}\frac{p'(y_i)}{p'(y)}\,R(y)\,W_1.\]

The determinant $A$ is less easy to compute. By using Laplace expansion, one can obtain that
\[A=(-1)^{\nu+1}\frac{(\nu+1)^{2\,\nu-4}}{\Delta^2}\,p'(y_i)\,\sum_{j=1}^\nu C_{ij}\,y^{\nu-j},\]
where $C_{ij}$ is the $(i, j)$-th cofactor of the matrix $U\cdot V$.

Let $U_{kl}$ and $V_{kl}$ be the $(k,l)$-th minor of the matrices $U, V$ respectively. Then the Binet--Cauchy formula implies that
\[C_{ij}=(-1)^{i+j}\sum_{k=1}^n U_{ik}\cdot V_{kj}.\]
The minors $U_{ik}$ are similar to the Vandermonde determinants,
\[U_{ik}=\frac{\prod_{1\le s<t\le n}(y_s-y_t)}{(-1)^{i-1}\,\prod_{s\ne i}(y_i-y_s)}e_{k-1}(\hat{y}_i),\]
where $e_{k}(\hat{y}_i)$ is the $k$-th elementary symmetric polynomial of $y_1, \dots, \hat{y}_i, \dots, y_n$.
Note that $y_1, \dots, y_n$ are roots of the polynomial $R(y)$, so these elementary symmetric polynomials
can be expressed as polynomials of $y_i$ and coefficients of $R(y)$.
It is also easy to compute the minors $V_{kj}$. Their explicit expressions are simple but not illuminating, so we omit them  here.

By using the above results, we obtain that
\[\p_iF=\frac1{(y-y_i)\,R'(y_i)}\frac{p'(y_i)}{p'(y)}\Big(R(y)-3\,F_y(x,y)\,\Sigma\Big),\]
where
\[\Sigma=\frac{y-y_i}{\Delta}\sum_{j=1}^\nu\sum_{k=1}^n (-1)^{j+1}\,e_{k-1}(\hat{y}_i)\,V_{kj}\,y^{\nu-j}.\]
When $\nu=3$, it is easy to show that $\Sigma=F_y(x_i,y)$. When $\nu=4$, after a very lengthy computation,
one can obtain that
\[\Sigma=F_y(x_i, y)+\frac{t_1^2}{5}(y-y_i)p'(y).\]
The lemma is proved.
\end{prf}

\begin{lem}\label{lem-17}
\begin{equation}
\Gamma_{ki}=3\,\frac{x_i+x_k}{(y_i-y_k)^2}\,\eta_{ii}, \label{eq-17}
\end{equation}
where $\eta_{ii}=-p'(y_i)/R'(y_i)$.
\end{lem}
\begin{prf}
One can prove the lemma directly by using Lemma \ref{lem-15}.
\end{prf}

\begin{rmk}
The vanishing of the G-functions of $E_6, E_8$ singularities can also be proved by the residue theorem, but the computation procedure becomes very long.
\end{rmk}

Although for $E_6, E_8$ we obtain the formula \eqref{eq-17} for the rotation coefficients, we still do not have a
simple way to relate the variables $y_i$ with $t_i$, as we did for the $A_n$ and $D_n$ cases. So at this moment
we can only check the validity of the conjecture for the $E_6, E_8$ singularities numerically. We first randomly
generate the complex values of  $t_1,\dots, t_n$, and solve the equations $F_x=0, F_y=0$ numerically to obtain the
values of the critical points $z^{(1)},\dots, z^{(n)}$.  Then one can determine the data $u^i, \, h_i,\, \gamma_{ij}$.
Our computation shows that Conjecture \ref{main-conj1} is valid in this numerical sense for the $E_6, E_8$ cases.

\vskip 1em

\begin{prfn}{Lemma \ref{easier-lem} for $E_6, E_8$ singularities}
First we have
\begin{align*}
&\sum_{1\le i < j\le n}\gamma_{ij}\frac{(h_i^2+h_j^2)^2}{h_i^3\,h_j^3}=
3\sum_{1\le i<j\le n}\frac{x_i+x_j}{(y_i-y_j)^2} \frac{(h_i^2+h_j^2)^2}{h_i^2\,h_j^2}\\
&=-3 \sum_{i=1}^n\sum_{j\ne i} \frac1{(y_i-y_j)^2}\left(\frac{q'(y_j) R'(y_i)}{p'(y_i) R'(y_j)}+
\frac{p'(y_j) q'(y_i)R'(y_i)}{p'(y_i)^2 R'(y_j)}+2\,\frac{q'(y_i)}{p'(y_i)}\right),
\end{align*}
then for fixed $i$,
\begin{align*}
-\sum_{j\ne i}\frac1{(y_i-y_j)^2}\frac{q'(y_j)}{R'(y_j)}&=\res_{y=y_i}\frac{q'(y)}{(y-y_i)^2\,R(y)},\\
-\sum_{j\ne i}\frac1{(y_i-y_j)^2}\frac{p'(y_j)}{R'(y_j)}&=\res_{y=y_i}\frac{p'(y)}{(y-y_i)^2\,R(y)},\\
-\sum_{j\ne i}\frac1{(y_i-y_j)^2}&=\lim_{y\to y_i}\frac{d}{dy}\left(\frac{R'(y)}{R(y)}-\frac1{y-y_i}\right),
\end{align*}
so we have
\begin{align*}
\sum_{1\le i < j\le n}&\gamma_{ij}\frac{(h_i^2+h_j^2)^2}{h_i^3\,h_j^3}
=3\sum_{i=1}^n \frac{R'(y_i)}{p'(y_i)}\res_{y=y_i}\frac{q'(y)}{(y-y_j)^2 R(y)}\\
&+3\sum_{i=1}^n \frac{q'(y_i) R'(y_i)}{p'(y_i)^2} \res_{y=y_i}\frac{p'(y)}{(y-y_j)^2 R(y)}\\
&+6\sum_{i=1}^n \frac{q'(y_i)}{p'(y_i)} \lim_{y\to y_i}\frac{d}{dy}\left(\frac{R'(y)}{R(y)}-\frac{1}{y-y_j}\right)\\
=&\sum_{i=1}^n \res_{y=y_i}g(y)=-\left(\res_{y=\infty}+\res_{y=\textrm{roots of}\ p'(y)}\right) g(y),
\end{align*}
where
\begin{align*}
g(y)=&\left(\frac32\,\frac{p'(y) q'''(y)+p'''(y) q'(y)}{p'(y)^2}\,\frac{R'(y)}{R(y)}\right.\\
&\left.-\frac32\frac{p'(y) q''(y)+p''(y) q'(y)}{p'(y)^2}\frac{R''(y)}{R(y)}+\frac{q'(y)\,R'''(y)}{p'(y)\,R(y)}\right).\\
\end{align*}
When $n=6$, $p'(y)$ has one root $y=-\frac{t_2}{2\,t_1}$. One can derive that
\[\res_{y=\infty}g(y)=\frac{12}{t_1},\quad  \res_{y=-\frac{t_2}{2 t_1}} g(y)=-\frac{12}{t_1},\]
so the $n=6$ case is proved.
When $n=8$, denote $a_1, a_2$ the two roots of $p'(y)$. We have
\[\res_{y=a_1} g(y)=-\res_{y=a_2} g(y)
=\frac{8(10 t_2 t_3+9 t_1 t_2 t_5-9 t_1^2 t_6)}{9 t_1^3 (a_1-a_2)^3},\]
and $\res_{y=\infty}g(y)=0$, so the $n=8$ case is also proved.
\end{prfn}

\subsection{The $E_7$ singularity} \label{sec-34}

In  this case $m=2$ and 
\[f(x,y)=x^3+x y^3.\]
The miniversal deformation can be chosen in the form
\[F(x,y)=x^3+p(y)\,x^2+q(y)\,x+r(y), \mbox{ where }\]
\[p(y)=t_1\,y+t_2,\ q(y)=y^3+t_3\,y+t_4,\ r(y)=t_5\,y^2+t_6\,y+t_7.\]
The critical points are determined from the following equations
\begin{align*}
F_x=&3\,x^2+2\,p(y)\,x+q(y)=0,\\
F_y=&p'(y)\,x^2+q'(y)\,x+r'(y)=0,
\end{align*}
which imply $x=\frac{Q}{P}$, and $R(y)=R(F_x,F_y,x)=Q^2-P\,S=0$, here
\[P=2\,p\,p'-3\,q',\ Q=3\,r'-p'\,q,\ S=q\,q'-2\,p\,r'.\]

\begin{lem}\label{lem-18}
We have
\begin{align}
\p_iF=&\frac{1}{(y-y_i)R'(y_i)}\frac{P(y_i)}{P(y)}\Big(R(y)-(P(y)\,x_i-Q(y))(P(y)\,x-Q(y))\Big)\nn\\
&+\frac{P(y_i)}{R'(y_i)}\Big(3(y+y_i)F_x-\frac{5\,t_1}{3}F_y\Big).
\end{align}
\end{lem}
The proof of the above lemma is very similar to the one of Lemma \ref{lem-15}, so we omit it.

By using the above lemma and eq. \eqref{eq-5}, one can prove the following
\begin{lem}\label{lem-19}
Let $\tilde{x}_k=x_k+\frac13\,p(y_k)$. Then
\begin{equation}
\Gamma_{ki}=3\frac{\tilde{x}_i+\tilde{x}_k}{(y_i-y_k)^2}\eta_{ii}
\end{equation}
where $\eta_{ii}=P(y_i)/R'(y_i)$.
\end{lem}

The above expression of $\Gamma_{ki}$ is similar to the one of the $E_8$ case. This fact has an interesting explanation.
Let us first introduce a modification of the miniversal deformation of the $E_7$ singularity
\[\tilde{F}=x^3+p(y)\,x^2+q(y)\,x+\tilde{r}(y),\mbox{ where } \tilde{r}(y)=r(y)+t_8\,y^5.\]
Make a coordinate transformation
\[\tilde{x}=x+\frac13\,p(y),\ \tilde{y}=\tau\left(y-\frac{t_1}{15\,t_8}\right),\mbox{ where } \tau=(t_8)^{\frac15}.\]
Then in these new coordinates the deformation $\tilde{F}$ reads 
\[\tilde{F}=\tilde{x}^3+\tilde{y}^5+\left(\tilde{t}_1\,\tilde{y}^3+\tilde{t}_2\,\tilde{y}^2+\tilde{t}_3\,\tilde{y}+\tilde{t}_4\right)\,\tilde{x}
+\tilde{t}_5\,\tilde{y}^3+\tilde{t}_6\,\tilde{y}^2+\tilde{t}_7\,\tilde{y}+\tilde{t}_8,\]
which is a miniversal deformation of an $E_8$-type singularity. Here $\tilde{t}_i\ (i=1, \dots, 8)$ are certain rational functions of $t_i\ (i=1, \dots, 7)$ and $\tau$, we omit their explicit expressions here.

Now let us take the limit $\tau\to0$. Then it is easy to see that one of the canonical coordinates, say $u^8$, goes to $\infty$, and the other seven ones become the canonical coordinates of the original $E_7$ singularity. By comparing Lemma \ref{lem-17} and \ref{lem-19}, one can also prove that the Christoffel symbols $\tilde{\Gamma}_{ki}$ associated to the $E_8$ singularity also tend to the Christoffel symbols $\Gamma_{ki}$ associated to the $E_7$ singularity, whenever $k, i=1, \dots, 7$.

By using the above observation, it is easy to see that if the G-function of the $E_8$ singularity vanishes, so does the G-function of the $E_7$ singularity. Similarly, if Lemma \ref{easier-lem} had been proved for the $E_8$ singularity, it also holds true for the $E_7$ singularity.

\subsection{The $\PP^1$-orbifold of $\tilde{A}_{p,q}$ type} \label{sec-35}

In this case, $m=3$, $(p,q,r)=(p,q,1)$, so $n=p+q$.  The tri-polynomial $F(z,t)$ reads
\[F(z,t)=-z^1z^2z^3+P_1(z^1)+P_2(z^2)+t_{n-1}+t_nz^3,\]
Its critical points are determined from the following equations
\begin{align}
&\p_{z^1}F=-z^2z^3+P'_1(z^1),\label{eq-a-Fx}\\
&\p_{z^2}F=-z^1z^3+P'_2(z^2),\label{eq-a-Fy}\\
&\p_{z^3}F=-z^1z^2+t_n.\label{eq-a-Fz}
\end{align}
We introduce an auxiliary function
\[\lambda(z)=P_1(z)+P_2(\frac{t^n}{z})+t_{n-1},\]
and denote $z_1, \dots, z_n$ its critical points. It is easy to see that $z_i$ coincides with the first component of the
critical point $z^{(i)}$ of $F(z,t)$, and the critical values of $\lambda(z)$ also coincide with the critical values of $F$,
so we have $u^i=\lambda(z_i)$.

The Hessian for $F$ reads
\begin{align*}
H=&P''_1(z^1)P''_2(z^2)P''_3(z^3)-2z^1z^2z^3\\
&\qquad -(z^1)^2P''_1(z^1)-(z^2)^2P''_2(z^2)-(z^3)^2P''_3(z^3).
\end{align*}
Then by using \eqref{eq-a-Fx}--\eqref{eq-a-Fz}, one obtains
\[\eta^{ii}=H(z^{(i)}(t),t)=-z_i^2\lambda''(z_i).\]

\begin{lem}\label{lem-a2}
\begin{equation}\label{eq-a-Fi}
\p_i\lambda(z)=\frac{z\lambda'(z)}{z_i(z-z_i)\lambda''(z_i)}.
\end{equation}
\end{lem}
\begin{prf}
Introduce
\[R(z)=\prod_{i=1}^n(z-z_i)=\frac{z^{q+1}}{p}\lambda'(z).\]
By using eq. \eqref{eq-2} and the Lagrange interpolation formula, one obtains
\[\p_i\left(z^q\lambda(z)\right)=z_i^q\frac{R(z)}{(z-z_i)R'(z_i)},\]
which implies the formula \eqref{eq-a-Fi} immediately. So we proved the lemma.
\end{prf}

\begin{lem}\label{lem-a4}
\begin{equation}
\Gamma_{ki}=\frac{z_k}{(z_k-z_i)^2z_i\lambda''(z_i)}.
\end{equation}
\end{lem}
\begin{prf}
The proof of this lemma is very similar to the derivation of eq. \eqref{eq-5}. We omit the details here.
\end{prf}

By using the above lemma we can verify, as we did in the $A_n$ and $D_n$ singularity cases, the 
Conjecture  \ref{main-conj1} for $n\le 8$ in the present case.

\begin{lem}\label{lem-a5}
\begin{equation}\label{G-func-A}
G(t)=-\frac{\log t_n}{24}.
\end{equation}
\end{lem}
\begin{prf}
By using the residue theorem, one obtains that
\[\p_iG=\frac{\eta_{ii}}{24}.\]
On the other hand, comparison of the coefficients of $z^{-q}$ in $\lambda(z)$ and $\p_i\lambda(z)$ yields
\[\p_i \log t_n =-\eta_{ii}.\]
The lemma is proved.
\end{prf}

\begin{lem}\label{lem-aoo}
\begin{equation}
O_1-O_2=\frac16(p^3+q^3-p-q).
\end{equation}
\end{lem}
\begin{prf}
Note that
\[h_i^{-2}=-z_i^2\lambda''(z_i),\ \gamma_{ij}=-\frac{h_ih_jz_iz_j}{(z_i-z_j)^2},\]
so one can prove the lemma by using the residue theorem.
\end{prf}

\subsection{The $\PP^1$-orbifold of $\tilde{D}_{r+2}$ type} \label{sec-36}

In this case, $m=3$, $(p,q,r)=(2,2,r)$, so $n=r+3$.  The tri-polynomial $F(z,t)$ reads
\[F(z,t)=-z^1z^2z^3+(z^1)^2+t_1z^1+(z^2)^2+t_2z^2+P_3(z^3),\]
Its critical points are determined from the following equations
\begin{align*}
&\p_{z^1}F=-z^2z^3+2z^1+t_1\\
&\p_{z^2}F=-z^1z^3+2z^2+t_2\\
&\p_{z^3}F=-z^1z^2+P_3'(z^3).
\end{align*}
Introduce an auxiliary function
\[\lambda(z)=P_3(z)+\frac{t_1^2+z\,t_1\,t_2+t_2^2}{z^2-4}.\]
and denote $z_1, \dots, z_n$ its critical points. Similarly to the $\tilde{A}_{p,q}$ cases, we have $u^i=\lambda(z_i)$.

The following lemmas are similar to the ones for the $\tilde{A}_{p,q}$ cases, so we omit their proofs.
It allows us to verify the Conjecture \ref{main-conj1} for $n\le 8$ in the present case.

\begin{lem}\label{lem-d2}
\begin{align*}
\eta^{ii}&=(4-z_i^2)\lambda''(z_i),\\
\p_i\lambda(z)&=\frac{4-z^2}{4-z_i^2}\frac{\lambda'(z)}{(z-z_i)\lambda''(z_i)},\\
\Gamma_{ki}&=\frac{4-z_kz_i}{\eta^{ii}(z_k-z_i)^2}.
\end{align*}
\end{lem}

\begin{lem}\label{lem-dg}
\begin{equation}\label{G-func-D}
G(t)=-\frac{\log t_n}{24\,r}
\end{equation}
\end{lem}

\begin{lem}\label{lem-doo}
\begin{equation}
O_1-O_2=\frac16(r^3-r)+2.
\end{equation}
\end{lem}

According to the results of  Lemma \ref{lem-a5}, \ref{lem-aoo}, \ref{lem-dg}, \ref{lem-doo},
we have the following conjecture.

\begin{conj}
For $\PP^1$-orbifolds of ADE type, we have
\begin{equation}
G(t)=-\frac{\log t_n}{24\,r},\ O_1-O_2=\frac16(p^3+q^3+r^3-p-q-r).
\end{equation}
\end{conj}

For $\PP^1$-orbifolds of $E$ type, we were unable to verify validity of the conjectures, even numerically, because
the numerical errors are too large in these cases.

\subsection{Some other examples} \label{sec-37}

\noindent{\bf{Example 3.26}}\ 
If the dimension of the Frobenius manifold equals 2, then it is easy to see that
\[O_1-O_2=\gamma_{12}\frac{(h_1^2+h_2^2)^2}{h_1^3h_2^3}=0,\]
since $h_1^2+h_2^2=0$. By using the formulae
\[h_1=\sqrt{-1}\,h_2,\quad \gamma_{12}=- \frac{\sqrt{-1}\, \mu_1}{u_1-u_2},\]
one can easily prove the following

\begin{lem}
The genus two G-function vanishes if and only if
\[\mu_1=\frac12,\ \frac13,\ \frac16,\]
which correspond to $A_1\times A_1$, $A_2$ and $\tilde{A}_{1,1}$ respectively.
\end{lem}

Note that the above three cases are also the only cases such that the genus one G-function $G(t)$ is
analytic on the caustics. 

\vskip 0.39cm 

\noindent{\bf{Example 3.26}}\ 
 Let $M$ be the Frobenius manifold corresponding to the quantum cohomology of $\PP^n\ (n\ge2)$. Then
$G^{(2)}(u,u_x,u_{xx})\ne 0$.

Indeed, the restrictions of the  $Q_p$ terms onto the small phase space vanish, while the restriction of $\F_2$ on the small phase space does not vanish in general. More generally, we obtain the following criterion.
\begin{lem}
The restriction of $\F_2$ on the small phase space vanishes if and only if $G^{(2)}|_{u^i_x=1, u^i_{xx}=0,\ 1\le i\le n}$.
\end{lem}
Since $\PP^n$ has non-trivial genus two Gromov--Witten invariants, so in this case $G^{(2)}(u,u_x,u_{xx})\ne 0$. 
 
\section{Conclusion} \label{sec-4}

It would be interesting to elucidate the geometric meaning of the genus two $G$-function $G^{(2)}$. In particular, the conditions for the vanishing of $G^{(2)}\equiv 0$ are of interest. Last but not least, finding of a higher genus $g\geq 3$ generalization of the decomposition \eqref{mthm-zh} is the main challenge. We plan to address these problems in a subsequent publication.

\vskip0.6 truecm
\noindent{\bf Acknowledgments.}
\vskip0.3 truecm

This work is partially supported by the European Research Council Advanced Grant FroM-PDE, by the
Russian Federation Government Grant No. 2010-220-01-077, by PRIN 2008 Grant ``Geometric methods
in the theory of nonlinear waves and their applications'' of Italian Ministry of Universities and Researches, and by the Marie Curie IRSES project RIMMP.
The work of Liu and Zhang  is also supported by the NSFC No.\,11071135 and No.\,11171176.

\appendix{}
\setcounter{section}{0}

\section{The genus two $G$-function}\label{app1}
 
The genus two $G$-function $G^{(2)}(u, u_x, u_{xx})$ depends rationally on the $x$-jets of the canonical coordinates
\begin{align}\label{g2-zh}
G^{(2)}(u, u_x, u_{xx})=&\sum_{i=1}^n G^{(2)}_i(u,u_x)u^i_{xx}+\sum_{i\ne j} G^{(2)}_{ij}(u) \frac{(u^j_x)^3}{u^i_x}\nn\\
&+ {\frac12\sum_{i,j}P^{(2)}_{ij}(u) u^i_x u^j_x}+\sum_{i=1}^n Q_i^{(2)} (u) \left(u_x^i\right)^2
\end{align}
with coefficients written in terms of the Lam\'e coefficients $h_i=h_i(u)$ and rotation coefficients $\gamma_{ij}=\gamma_{ij}(u)$ of the semisimple Frobenius manifold. To simplify the expressions of these coefficients, we use the function
\[H_i=\frac12\sum_{j\ne i} u_{ij}\gamma_{i j}^2,\quad 1\le i\le n\]
with $u_{ij}=u_i-u_j$, these functions are given by the gradients of the isomonodromic tau function of the Frobenius manifold \cite{DZ-cmp, DZ-select}. Then we have
\begin{align}
G^{(2)}_i=&\frac{\p_x h_i\,H_i}{60\, u_{i,x} h_i^3}-\frac{3\,\p_i h_i H_i}{40\,h_i^3}
+\frac{19\,(\p_i h_i)^2}{2880\,h_i^4}-\frac{7\,\p_i h_i \p_x h_i}{5760\,u_{i,x} h_i^4}\nn
\\
&+\sum _{k} \left[\frac{\gamma_{i k} H_i}{120\, h_i h_k}+
\frac{\gamma_{i k} H_k}{120\, h_i h_k}\left(7+\frac{u_{k,x}}{u_{i,x}}\right) 
-\frac{\gamma_{i k}}{5760\,h_i^2 h_k}\left(4\,\p_i h_i+\frac{\p_x h_i}{u_{i,x}}\right)
\right.\nn\\
&\left.
-\frac{\gamma_{i k}\p_k h_k}{h_i h_k^2}\left(\frac{u_{k,x}}{1152\,u_{i,x}}+
\frac{7}{2880}\right)+\frac{\gamma_{i k} \p_k h_k}{384\,h_i^3}
-\frac{\p_k \gamma_{i k} h_k}{384\,h_i^3}
+\frac{\p_i \gamma_{i k}h_k u_{k,x}}{1920\,u_{i,x}h_i^3}\right.\nn\\
&\left.
+\frac{\p_i\gamma_{i k}}{2880\,h_i h_k}
+\frac{\p_x\gamma_{i k}}{5760\,u_{i,x}h_i h_k}
+\frac{\p_k\gamma_{i k}}{h_i h_k}\left(\frac{u_x^k}{2880\, u_x^i}+\frac{7}{2880}\right)+\frac{\gamma_{i k} h_i \p_k h_k}{2880\,h_k^4}
\right.\nn\\
&\left.-\frac{\gamma_{ik}^2}{h_i^2}\left(\frac{7 u_x^k}{1152\, u_x^i}+\frac{19}{720}\right)+\frac{\gamma_{ik}^2}{1440\, h_k^2}\right]-\sum _{k, l} \left(\frac{h_i \gamma_{il} \gamma_{kl}}{2880\, h_k h_l^2}+\frac{u_{k,x}h_k \gamma_{il} \gamma_{kl}}{1920\, u_x^i h_i h_l^2}\right),\nn
\end{align}

\begin{align*}
G^{(2)}_{ij}=&
-\frac{\gamma_{i j}^2 H_j}{120\, h_j^2}
+\frac{\gamma _{i j}^3}{480\, h_i h_j}
-\frac{\gamma _{i j}}{5760}\left(\frac{\p_i\gamma_{i j}}{h_i^2}+\frac{\p_j\gamma_{i j}}{h_j^2}\right)
+\frac{\gamma _{i j}^2}{5760}\left(\frac{\p_i h_{i}}{h_i^3}+\frac{3\,\p_j h_{j}}{h_j^3}
\right)\\
&+\sum_k \left(
\frac{\gamma_{i j}\gamma_{i k}\gamma_{j k}}{5760\, h_k^2}+\frac{\gamma_{i j}^2}{5760\,h_k}\left(\frac{\gamma_{j k}}{h_j}-\frac{\gamma_{i k}}{h_i}\right)
\right),
\end{align*}

\begin{align*}
P^{(2)}_{ij}&=-\frac{2\, \gamma_{i j} H_i H_j}{5\,h_i h_j}
+\frac{\gamma_{i j} \p_j h_j H_i}{20\, h_i h_j^2}
+\frac{\gamma_{i j} h_i \p_j h_j H_j}{20\,h_j^4}
-\frac{19\,\gamma_{i j}^2 H_j}{30\,h_j^2}
-\frac{\p_i\gamma_{i j} H_j}{60\,h_i h_j}\\
&
+\frac{41\,\gamma_{i j}^3}{240\,h_i h_j}
-\frac{41 \gamma _{i j} \partial_i\gamma_{ij}}{1440\, h_i^2 }
+\frac{\p_i\gamma_{i j}\p_j h_j}{1440\,h_i h_j^2}
+\frac{79\,\gamma_{i j}^2 \p_j h_j}{1440\,h_j^3}
-\frac{\gamma_{i j} \p_i h_i \p_j h_j}{720\, h_i^2 h_j^2}
-\frac{\gamma_{i j} h_i (\p_j h_j)^2}{288\,h_j^5}\\
&+\sum_k\left(
\frac{\gamma_{i j}\gamma_{i k} H_j}{60\,h_j h_k}
-\frac{\gamma_{i k}\gamma_{j k} h_i h_j H_k}{30\, h_k^4}
-\frac{\gamma_{i j}\gamma_{j k} h_i H_j}{60\, h_j^2 h_k}
+\frac{\gamma_{i k}\gamma_{j k} h_i H_j}{60\, h_j h_k^2}
-\frac{7\,\gamma_{i j}\gamma_{j k} h_i H_k}{60\, h_j^2 h_k}\right.\\
&\qquad\left.
-\frac{\gamma_{i j}\gamma_{i k} \p_j h_j}{720\, h_j^2 h_k}
+\frac{\gamma_{i j}\gamma_{j k} h_i  \p_j h_j}{240\, h_j^3 h_k}
-\frac{\gamma_{i k}\gamma_{j k} h_i  \p_j h_j}{1440\, h_j^2 h_k^2}
+\frac{\gamma_{i j}\gamma_{j k} h_i  \p_k h_k}{720\, h_k^4}
+\frac{\gamma_{i k}\gamma_{j k} h_i h_j  \p_k h_k}{288\, h_k^5}
\right.\\
&\qquad\left.+\frac{ \gamma _{j k}\,\partial_i\gamma _{i j}}{1440\, h_i h_k}
-\frac{h_j h_k \gamma _{i j} \partial_i\gamma _{i k}}{360\, h_i^4 }
-\frac{h_j (3\,\gamma_{ik}\partial_i\gamma_{ij}+2\,\gamma_{i j} \p_i\gamma_{i k})}{1440\, h_i^2 h_k }
-\frac{7\,h_j \gamma _{i j} \partial_k (h_k^{-1}\gamma _{i k})}{1440\, h_i^2 }
\right.\\
&\qquad\left.
-\frac{h_i h_j \gamma _{i k}\, \partial_k\gamma _{j k}}{480 \,h_k^4 }+\frac{\gamma _{i j}^2 \gamma _{j k}}{120\, h_j h_k}
+\frac{7\, h_i \gamma _{i j} \gamma _{j k}^2}{160 \,h_j^3}
+\frac{11 \gamma _{i j} \gamma _{i k} \gamma _{j k}}{2880\, h_k^2}+\frac{h_j \gamma _{i k}^2 \gamma _{j k}}{96 \,h_k^3}\right)\\
&+\sum_{k, l}\left(
\frac{h_i h_j \gamma _{i l} \gamma _{j l} }{720\, h_k h_l^2}
\left(\frac{\gamma _{k l}}{h_l}-\frac{\gamma_{j k}}{2 h_j}\right)
-\frac{h_i  \gamma_{i j}\gamma _{j l} \gamma _{k l} }{720\,h_k h_l^2}
\right),
\end{align*}

\begin{align*}
Q^{(2)}_{i}
=&\frac{4 H_i^3}{5\, h_i^2}
-\frac{7\, \p_i h_i H_i^2}{10\,h_i^3}
+\frac{7\, (\p_i h_i)^2 H_i}{48\, h_i^4}-\frac{(\p_i h_i)^3}{120\,h_i^5}+\sum_k \left(\frac{7 \gamma_{i k} H_i H_k }{10\, h_i h_k}
-\frac{\gamma_{i k}\p_i h_i H_i}{120\, h_i^2 h_k}\right.\\
&\left.
+\frac{7 \,\p_k\left(h_k^{-1} \gamma_{i k}\right)H_i}{240\, h_i }
-\frac{7  \gamma_{i k} \p_i h_i H_k}{80\, h_i^2 h_k}
+\frac{\gamma_{i k}H_k}{576\, u_{i k} h_i h_k}
+\frac{(2 H_i+7 H_k)\p_i\gamma_{i k}}{240\,h_i h_k}\right.\\
&\left.
+\frac{\gamma_{i k} h_k H_i}{576\, u_{i k} h_i^3}
-\frac{31  \gamma_{i k}^2 H_i}{144\, h_i^2}
+\frac{\gamma_{i k} (\p_i h_i)^2}{720\, h_i^3 h_k} 
+\frac{253\,\gamma_{i k}^2\p_i h_i }{5760\,h_i^3}
-\frac{\p_i\gamma_{i k} \p_i h_i }{960\,h_i^2 h_k}
-\frac{\gamma_{i k}^2 \p_k h_k}{2880 \,h_k^3}\right.\\
&\left.-\frac{7\, \p_k\left(h_k^{-1}\gamma_{i k}\right)\p_i h_i}{1920\, h_i^2 }
-\frac{7\,\p_i\gamma_{i k}\p_k h_k}{5760\,h_i h_k^2}
-\frac{41\,\p_i\gamma_{i k} \p_i h_i h_k}{5760\,h_i^4}
+\frac{ \p_i (h_i\gamma_{i k}) \p_k h_k}{2880\, h_k^4 }\right.\\
&\left.
-\frac{113\, \gamma _{i k}\partial_i\gamma_{i k}}{5760\, h_i^2}
+\frac{\left(3\,\partial_i\gamma_{i k}+\partial_k\gamma_{i k}\right) \gamma _{i k}}{1440\, h_k^2 }
-\frac{ \partial_i \gamma _{i k}h_k}{576\, u_{i k} h_i^3 }
-\frac{\partial_k \gamma _{i k}}{576\, u_{i k} h_i h_k }
-\frac{\gamma _{i k}^3}{240\, h_i h_k}\right)\\
&+\sum_{k, l}  \left(-\frac{\gamma_{k l}\partial_i (h_i \gamma _{i l})}{2880\, h_k h_l^2 }
+\frac{\gamma _{i l}^2 \gamma _{k l}}{2880 \,h_k h_l}
-\frac{\gamma _{i k} \gamma _{i l}^2}{240\,h_i h_k }
-\frac{\gamma_{k l}\partial_i\gamma _{i k} }{2880 \,h_i h_l }
+\frac{u_{l k}\gamma _{i k}\partial_l \gamma _{k l} }{1152\,u_{i l} h_i h_l }
\right.\\
&\left.
+\frac{ u_{k l}\gamma _{i k} \gamma_{k l} \partial_i\gamma _{i l} }{144\,  h_i^2}
+\frac{ h_l \gamma _{i k}\partial_i \gamma _{i l}}{1440 \, h_i^2 h_k}
+\frac{h_k u_{k l} \gamma_{k l} \partial_i \gamma _{i l} }{1152\, u_{i k} h_i^3}
+\frac{h_l u_{i k}\gamma _{i k}^2 \partial_i\gamma _{i l}}{40\,h_i^3}
\right).
\end{align*}

In these expressions, the summations are taken over indices such that the denominators do not vanish.

\section{General formula for the genus two free energy}\label{app2}
In this formula derived in \cite{DZ} the following notations are used
$$
V_{ij}=(u_j-u_i) \gamma_{ij}, \quad u_{ij}=u_i-u_j.
$$ 
A summation over repeated indices is assumed in each term
of the formula provided the denominators do not vanish.
\begin{eqnarray}
&&
{\cal F}_2=\frac{1}{1152}\frac{u_i^{IV}}{{u_i'}^2\,h_i^2}
-\frac{7}{1920}\,\frac{{u_i''}\,u_i'''}{{u_i'}^3\,h_i^2}+
\frac{1}{360}\frac{{u_i''}^3}
{{u_i'}^4\,h_i^2}
+\frac1{40}\frac{V_{ij}^2\,u_i'''}{u_{ij}\,{u_i'}\,h_i^2}\nn\\
&&+\frac1{640}\frac{V_{ij}\,h_j\,{u_j'}\,u_i'''}
{u_{ij}\,{u_i'}^2\,h_i^3}-
\frac{19}{2880}\frac{V_{ij}\,u_i'''\,h_j}{u_{ij}\,{u_i'}\,
h_i^3}+
\frac1{1152}\frac{V_{ij}\,u_i'''\,h_i}{u_{ij}\,{u_j'}\,
h_j^3}+\frac7{40}\frac{V_{ij}^2\,V_{ik}^2\,{u_i''}}{u_{ij}\,u_{ik}\,h_i^2}\nn\\
&&
-\frac1{240}\frac{V_{ij}^2\,V_{ik}\,{u_i''}\,h_k
\left(32\,{u_i'}-7\,{u_k'}\right)}
{u_{ij}\,u_{ik}\,{u_i'}\,h_i^3}+\frac{1}{40}\frac{V_{ij}\,V_{jk}^2\,{u_i''}\,h_i}
{u_{ij}\,u_{jk}\,h_j^3}-
\frac{1}{48}\frac{V_{ij}\,V_{jk}^2\,{u_j'}\,{u_i''}}
{u_{ij}\,u_{jk}\,{u_i'}\,h_i\,h_j}\nn\\
&&-
\frac{3}{64}\frac{V_{ij}^2\,{u_i''}}{u_{ij}^2\,h_i^2}-\frac{11}{480}\frac{V_{ij}^2\,{u_i''}^2}{u_{ij}\,{u_i'}^2\,
h_i^2}+
\frac{29}{5760}\frac{V_{ij}\,V_{jk}\,{u_i''}\,h_i\,h_k
\left({u_k'}-2\,{u_j'}\right)}
{u_{ij}\,u_{jk}\,{u_j'}\,h_j^4}\nn\\
&&+
\frac{1}{384}\frac{V_{ij}\,V_{ik}\,{u_i''}\,h_k
\left({u_i'}-{u_k'}\right)}
{u_{ij}\,u_{ik}\,{u_j'}\,h_j^3}+
\frac{1}{1920}\frac{V_{ij}\,V_{ik}\,{u_i''}\,h_j\,h_k
\left(54\,{u_i'}^2-25\,{u_i'}\,{u_j'}-{u_j'}\,{u_k'}\right)}
{u_{ij}\,u_{ik}\,{u_i'}^2\,h_i^4}\nn\\
&&+
\frac{1}{576}\frac{V_{ij}\,V_{jk}\,{u_i''}\,h_k
\left(2\,{u_j'}-{u_k'}\right)}
{u_{ij}\,u_{jk}\,{u_i'}\,h_i\,h_j^2}-
\frac{1}{5760}\frac{V_{ij}\,V_{jk}\,{u_k'}\,{u_i''}\,h_k
\left(27\,{u_i'}+{u_k'}\right)}
{u_{jk}\,u_{ik}\,{u_i'}^2\,h_i^3}\nn\\
&&-
\frac{19}{1920}\frac{V_{ij}\,V_{jk}\,{u_i''}\,h_k}
{u_{ij}\,u_{ik}\,h_i^3}+
\frac{1}{5760}\frac{V_{ij}\,V_{jk}\,h_k
\left(27\,{u_i'}\,{u_k'}-{u_j'}^2+2\,{u_j'}\,{u_k'}\right)\,{u_i''}}
{u_{ij}\,u_{jk}\,{u_i'}^2\,h_i^3}\nn\\
&&+
\frac{1}{288}\frac{V_{ij}\,V_{jk}\,{u_i''}\,h_i}
{u_{jk}\,u_{ik}\,h_k^3}+
\frac{1}{384}\frac{V_{ij}\,V_{jk}\,{u_i'}\,{u_i''}\,h_i}
{u_{ij}\,u_{ik}\,{u_k'}\,h_k^3}-
\frac{1}{576}\frac{V_{ij}\,V_{jk}\,{u_k'}\,{u_i''}}
{u_{jk}\,u_{ik}\,{u_i'}\,h_i\,h_k}\nn\\
&&-
\frac{1}{384}\frac{V_{ik}\,V_{jk}\,{u_k'}\,{u_i''}\,h_i}
{u_{ik}\,u_{jk}\,{u_j'}\,h_j^3}+
\frac{1}{1920}\frac{V_{ij}\,{u_i''}^2\,h_j
\left(11\,{u_i'}-5\,{u_j'}\right)}
{u_{ij}\,{u_i'}^3\,h_i^3}-
\frac{1}{5760}\frac{V_{ij}\,{u_i''}\,{u_j''}\,h_j}
{u_{ij}\,{u_i'}^2\,h_i^3}\nn\\
&&+
\frac{1}{5760}\frac{V_{ij}\,{u_i''}\,h_j
\left(57\,{u_i'}^2-27\,{u_i'}\,{u_j'}-{u_j'}^2\right)}
{u_{ij}^2\, {u_i'}^2\,h_i^3}+\frac{1}{1152}\frac{V_{ij}\,{u_i''}\,h_i
\left(4\,{u_j'}-3\,{u_i'}\right)}
{u_{ij}^2\, {u_j'}\,h_j^3}\nn\\
&&\cla{}\nn\\
&&-
\frac{1}{576}\frac{V_{ij}\,{u_j'}\,{u_i''}}
{u_{ij}^2\,{u_i'}\,h_i\,h_j}-
\frac{1}{1152}\frac{V_{ij}\,{u_i''}\,{u_j''}}
{u_{ij}\,{u_i'}\,{u_j'}\,h_i\,h_j}+
\frac1{10}\frac{V^2_{ij}\,V^2_{ik}\,V^2_{il}\,{u_i'}^2}
{u_{ij}\,u_{ik}\,u_{il}\,h_i^2}\nn\\
&&
-\frac7{20}\frac{V^2_{ij}\, V^2_{ik}\, V_{il}\,h_l\,{u_i'}^2}
{u_{ij} u_{ik} u_{il}\,h_i^3}
+\frac7{40}\frac{V^2_{ij}\, V^2_{ik}\, V_{il}\,
h_l\,{u_i'}\,{u_l'}}
{u_{ij}\, u_{ik}\, u_{il}\, h_i^3}
-\frac1{8} \frac{V^2_{ij}\,V_{ik}\,V^2_{kl}\,{u_i'}\,{u_k'}}
{u_{ij}\,u_{ik}\,u_{kl}\,h_i\,h_k}\nn\\
&&+
\frac1{40} \frac{V^2_{ij}\,V_{ik}\,V_{kl}\,h_l
\left({u_k'}^2-3\,{u_i'}^2-2\,{u_k'}\,{u_l'}\right)}
{u_{ij}\,u_{ik}\,u_{kl}\,h_i^3}
+\frac3{40}\frac{V^2_{ij}\,V_{ik}\,V_{kl}\,{u_i'}\,{u_l'}\,h_l}
{u_{ij}\,u_{ik}\,u_{il}\,h_i^3}\nn\\
&&
+\frac1{40}\frac{V^2_{ij}\,V_{ik}\,V_{kl}\,h_l\left(
3\,{u_i'}^2+{u_l'}^2\right)}
{u_{ij}\,u_{kl}\,u_{il}\,h_i^3}
+\frac1{48}\frac{V^2_{ij}\,V_{ik}\,V_{kl}\,h_l\,{u_i'}
\left(2\,{u_k'}-{u_l'}\right)}{u_{ij}\,u_{ik}\,u_{kl}\,h_i\,
h_k^2}\nn\\
&&+\frac5{96} \frac{V^2_{ij}\,V_{ik}\,V_{il}\,h_k\,h_l\,
\left(4\,{u_i'}^2-4\,{u_i'}\,{u_k'}+{u_k'} {u_l'}\right)}
{u_{ij} u_{ik} u_{il}\,h_i^4}
-
\frac{83}{480}\frac{V^2_{ij}\,V^2_{ik}\,{u_i'}^2}
{u_{ij}\,u_{ik}^2\,h_i^2}\nn
\end{eqnarray}
\begin{eqnarray}
&&+
\frac1{144}\frac{V_{ij}\,V_{ik}\,V_{jl}\,V_{kl}\,{u_i'}^2}
{u_{ik}\,u_{jl}\,u_{il}\,h_i^2}
-\frac1{144}\frac{V_{ij}\,V_{ik}\,V_{jl}\,V_{kl}\,{u_i'}^2}
{u_{ij}\,u_{ik}\,u_{kl}\,h_i^2}-
\frac1{48}\frac{V_{ij}^2\,V_{ik}\,V_{kl}\,{u_i'}\,{u_l'}}
{u_{ij}\,u_{kl}\,u_{il}\,h_i\,h_l}
\nn\\
&&+\frac{29}{1920} \frac{V_{ij}\,V_{ik}\,V_{jl}\,h_k\,h_l
\left({u_k'}\,{u_l'}-{u_i'}\,{u_k'}+2\,{u_i'}^2-2\,{u_i'}\,{u_l'}\right)}
{u_{ij}\,u_{ik}\,u_{il}\,h_i^4}\nn\\
&&-
\frac{29}{5760} \frac{V_{ij}\,V_{ik}\,V_{jl}\,h_k\,h_l\,
{u_j'}\left(2\,{u_k'}\,{u_l'}+2\,{u_i'}\,{u_j'}-{u_j'}\,{u_k'}
-4\,{u_i'}\,{u_l'}\right)}
{u_{ij}\,u_{ik}\,u_{jl}\,h_i^4\,{u_i'}}\nn\\
&&-
\frac{1}{1152} \frac{V_{ij}\,V_{ik}\,V_{jl}\,h_k\,h_l\,
\left(4\,{u_i'}\,{u_j'}-4\,{u_i'}\,{u_l'}+{u_k'}\,{u_l'}\right)}
{u_{ij}\,u_{ik}\,u_{jl}\,h_i^2\,h_j^2}\nn\\
&&-\frac1{384}\frac{V_{ij}\,V_{ik}\,V_{jl}\,h_l
\left({u_i'}\,{u_j'}^2-2\,{u_j'}\,{u_i'}\,{u_l'}\right)}
{u_{ij}\,u_{ik}\,u_{jl}\,{u_k'}\,h_k^3}-\frac{29}{5760} \frac{V_{ij}\,V_{ik}\,V_{jl}\,h_k\,h_l\,
{u_l'}^2\left(2\,{u_i'}-{u_k'}\right)}
{u_{ik}\,u_{jl}\,u_{il}\,h_i^4\,{u_i'}}\nn\\
&&+
\frac1{1152}\frac{V_{ij}\,V_{ik}\,V_{jl}\,h_l\,{u_i'}^2
\left({u_i'}-3\,{u_l'}\right)}
{u_{ij}\,u_{ik}\,u_{il}\,{u_k'}\,h_k^3}
-
\frac1{384}\frac{V_{ij}\,V_{ik}\,V_{jl}\,h_l\,{u_i'}\,{u_l'}^2}
{u_{ik}\,u_{jl}\,u_{il}\,{u_k'}\,h_k^3}\nn\\
&&-
\frac1{1152}\frac{V_{ij}\,V_{ik}\,V_{jl}\,h_l\,{u_j'}^2
\left(3\,{u_l'}-2\,{u_j'}\right)}
{u_{ij}\,u_{jl}\,u_{jk}\,{u_k'}\,h_k^3}-
\frac1{288}\frac{V_{ij}\,V_{ik}\,V_{jl}\,h_l\,{u_j'}
\left({u_j'}-2\,{u_l'}\right)}
{u_{ik}\,u_{jl}\,u_{jk}\,h_k^3}\nn\\
&&+
\frac1{576}\frac{V_{ij}\,V_{ik}\,V_{jl}\,h_l\,{u_k'}
\left(2\,{u_k'}-3\,{u_l'}\right)}
{u_{ik}\,u_{jk}\,u_{kl}\,h_k^3}-
\frac1{1152}\frac{V_{ij}\,V_{ik}\,V_{jl}\,h_l\,{u_l'}^3}
{u_{jl}\,u_{kl}\,u_{il}\,{u_k'}\,h_k^3}\nn\\
&&+
\frac1{288}\frac{V_{ij}\,V_{ik}\,V_{jl}\,h_l\,{u_l'}^2}
{u_{ik}\,u_{jl}\,u_{kl}\,h_k^3}-
\frac1{576}\frac{V_{ij}\,V_{ik}\,V_{jl}\,h_k\,{u_l'}
\left({u_k'}-2\,{u_i'}\right)}
{u_{ik}\,u_{jl}\,u_{il}\,h_i^2\,h_l}\nn\\
&&-
\frac7{1440}\frac{V_{ij}\,V_{ik}\,V_{il}\,h_j\,h_k\,
h_l\left(8\,{u_i'}^3-12\,{u_i'}^2\,{u_j'}-
{u_j'}\,{u_k'}\,{u_l'}+6\,{u_i'}\,{u_j'}\,{u_k'}\right)}
{u_{ij}\,u_{ik}\,u_{il}\,h_i^5\,{u_i'}}\nn\\
&&-
\frac1{1152}\frac{V_{ij}\,V_{ik}\,V_{jl}\,{u_k'}\,{u_l'}}
{u_{ik}\,u_{jl}\,u_{kl}\,h_k\,h_l}-
\frac{29}{1152}\frac{V_{ij}\,V_{ik}\,V_{jk}\,{u_i'}^2}
{u_{ij}\,u_{ik}^2\,h_i^2}-
\frac{53}{1920}\frac{V_{ij}^2\,V_{ik}\,h_k\,{u_i'}\,{u_k'}}
{u_{ij}\,u_{ik}\,u_{jk}\,h_i^3}\nn\\
&&-
\frac{1}{320}\frac{V_{ij}^2\,V_{ik}\,h_k
\left(3\,{u_i'}^2-8\,{u_k'}^2\right)}
{u_{ij}\,u_{ik}^2\,h_i^3}-
\frac{V_{ij}^2\,V_{ik}\,{u_i'}\,h_k}{u_{ij}^2\,u_{jk}\,h_i^3}
\left(\frac{27}{640}\,{u_k'}-\frac{233}{2880}\,{u_i'}\right)\nn\\
&&-
\frac{V_{ij}^2\,V_{ik}\,{u_i'}\,h_k}
{u_{ik}^2\,u_{jk}\,h_i^3}
\left(\frac{233}{2880}\,{u_i'}-\frac{67}{960}\,{u_k'}\right)
+
\frac{1}{1152}\frac{V_{ij}^2\,V_{ik}\,h_i\,{u_i'}^3}
{u_{ij}\,u_{ik}^2\,{u_k'}\,h_k^3}-
\frac{1}{576}\frac{V_{ij}^2\,V_{ik}\,h_i\,{u_i'}^3}
{u_{ij}^2\,u_{ik}\,{u_k'}\,h_k^3}\nn\\
&&-
\frac{1}{48}\frac{V_{ij}^2\,V_{ik}\,{u_i'}\,{u_k'}}
{u_{ij}\,u_{ik}^2\,h_i\,h_k}+
\frac{233}{1440}\frac{V_{ij}^3\,h_j\,{u_i'}^2}
{u_{ij}^3\,h_i^3}-
\frac{43}{384}\frac{V_{ij}^3\,h_j\,{u_i'}\,{u_j'}}
{u_{ij}^3\,h_i^3}-
\frac{1}{12}\frac{V_{ij}^3\,{u_i'}\,{u_j'}}
{u_{ij}^3\,h_i\,h_j}\nn\\
&&+
\frac{29}{5760}\frac{V_{ij}\,V_{ik}\,
h_j\,h_k
\left(3\,{u_i'}\,{u_k'}+3\,{u_j'}\,{u_k'}+6\,{u_i'}\,{u_j'}
-6\,{u_i'}^2-2\,{u_j'}^2\right)}
{u_{ij}^2\,u_{ik}\,h_i^4}\nn\\
&&+
\frac{29}{5760}\frac{V_{ij}\,V_{ik}\,{u_j'}\,{u_k'}\,
h_j\,h_k
\left({u_k'}-6\,{u_i'}\right)}
{u_{ij}\,u_{ik}^2\,{u_i'}\,h_i^4}+
\frac{1}{576}\frac{V_{ij}\,V_{ik}\,{u_j'}\,
h_k\left(2\,{u_i'}-{u_k'}\right)}
{u_{ij}^2\,u_{ik}\,h_i^2\,h_j}
\nn\\
&&+
\frac{1}{1152}\frac{V_{ij}\,V_{ik}\,u_{ij}\,h_k
\left(3\,{u_i'}^2\,{u_k'}-3\,{u_i'}\,\,{u_k'}^2+
{u_k'}^3-{u_i'}^3\right)}
{u_{ik}^2\,u_{jk}^2\,{u_j'}\,h_j^3}\nn\\
&&+
\frac{1}{576}\frac{V_{ij}\,V_{ik}\,u_{ik}\,h_k
\left(-{u_i'}^3+3\,{u_j'}^2\,{u_k'}-4\,{u_i'}\,{u_j'}\,{u_k'}+
2\,{u_i'}^2\,{u_j'}-2\,{u_j'}^3\right)}
{u_{ij}^2\,u_{jk}^2\,{u_j'}\,h_j^3}\nn
\end{eqnarray}
\begin{eqnarray}
&&+
\frac{1}{384}\frac{V_{ij}\,V_{ik}\,h_k
\left(-{u_i'}\,{u_k'}^2+{u_i'}^3-6\,{u_j'}^2\,{u_k'}\right)}
{u_{ij}\,u_{jk}^2\,{u_j'}\,h_j^3}+
\frac{1}{384}\frac{V_{ij}\,V_{ik}\,h_k
\,{u_i'}^2\,{u_k'}}
{u_{ij}^2\,u_{jk}\,{u_j'}\,h_j^3}\nn\\
&&+
\frac{1}{288}\frac{V_{ij}\,V_{ik}\,h_k
\left(4\,{u_i'}\,{u_k'}+
{u_k'}^2-2\,{u_i'}^2+3\,{u_j'}^2\right)}
{u_{ij}\,u_{jk}^2\,h_j^3}-
\frac{1}{576}\frac{V_{ij}\,V_{ik}
\,{u_j'}\,{u_k'}}
{u_{ik}\,u_{jk}^2\,h_j\,h_k}\nn\\
&&+
\frac{1}{384}\frac{V_{ij}\,V_{ik}\,h_k
\left(2\,{u_i'}\,{u_k'}^2-
{u_i'}^2\,{u_k'}-{u_k'}^3\right)}
{u_{ik}\,u_{jk}^2\,{u_j'}\,h_j^3}+
\frac{1}{288}\frac{V_{ij}\,V_{ik}\,h_k
\left({u_k'}^2-2\,{u_i'}\,{u_k'}
+{u_i'}^2\right)}
{u_{ik}\,u_{jk}^2\,h_j^3}\nn\\
&&
\nn\\
&&+
\frac{1}{1152}\frac{V_{ij}^2\,{u_i'}
\left(37\,{u_i'}\,{u_j'}\,h_j^2+
10\,{u_i'}\,{u_j'}\,h_i^2-3\,{u_i'}^2\,h_i^2
+11\,{u_j'}^2\,h_j^2\right)}
{u_{ij}^3\,{u_j'}\,h_i^2\,h_j^2}\nn\\
&&-
\frac{1}{576}\frac{V_{ij}\,h_j
\left(4\,{u_i'}^3+
4\,{u_i'}\,{u_j'}^2-6\,{u_i'}^2\,{u_j'}-{u_j'}^3\right)}
{u_{ij}^3\,{u_i'}\,h_i^3}+
\frac{1}{576}\frac{V_{ij}\,{u_i'}\,{u_j'}}
{u_{ij}^3\,h_i\,h_j}.\nn
\end{eqnarray}


\begin{thebibliography}{99}
\bibitem{D1} B.\,Dubrovin,  Geometry of 2D topological field theories,
Integrable systems and quantum groups (Montecatini Terme, 1993), 120--348, Lecture Notes in Math., 1620, Springer, Berlin, 1996. 
\bibitem{dz-comp} B.\,Dubrovin, Y.\,Zhang, Extended affine Weyl groups and Frobenius manifolds, Compositio Mathematica {\bf 111} (1998) 167-219.
\bibitem{DZ-cmp} B.\,Dubrovin, Y.\,Zhang, Bi-Hamiltonian hierarchies in 2D TFT at one-loop approximation, Comm. Math. Phys. 198 (1998), no. 2, 311--361.
\bibitem{DZ-select} B.\,Dubrovin, Y.\,Zhang, Frobenius manifolds and Virasoro constraints, Sel. Math., New ser. 5 (1999), 423--466.
\bibitem{DZ} B.\,Dubrovin, Y.\,Zhang, 
Normal forms of integrable PDEs, Frobenius manifolds and Gromov--Witten invariants, arXiv: math/0108160
\bibitem{dlz} B.\,Dubrovin, S.-Q.\, Liu, Y.\,Zhang Frobenius Manifolds and Central Invariants for the Drinfeld - Sokolov Bihamiltonian Structures, Adv. Math.  {\bf 219} (2008) 780-837.
 \bibitem{htt} B.\,Dubrovin, S.-Q.\, Liu, Y.\,Zhang, Integrable hierarchies of topological type, to appear.
\bibitem{eguchi0} T.\,Eguchi, C.-S.\,Xiong, Quantum cohomology at higher genus: topological recursion relation, Adv. Theor. Math. Phys. {\bf 2} (1998) 219-229.
\bibitem{eguchi1} T.\,Eguchi, Y.\,Yamada, S.-K.\,Yang, On the genus expansion in the topological string theory, Rev. Math. Phys. {\bf 7} (1995) 279-309.
\bibitem{eguchi2} T.\,Eguchi, E.\,Getzler, C.-S.\,Xiong, Topological gravity in genus 2 with two primary fields, Adv. Theor. Math. Phys. {\bf 4} (2000) 981-998; Erratum - ibid. {\bf 5} (2001) 211-212.
\bibitem{fsz} C.\,Faber, S.\,Shadrin, D.\,Zvonkine, Tautological relations and the $r$-spin Witten conjecture, arXiv:math/0612510.
\bibitem{for-1}H.\,Fan, T.\,Jarvis, Y.\,Ruan, Geometry and analysis of spin equations, Comm. Pure Appl. Math. 61(2008), 745-788.
\bibitem{for} H.\,Fan, T.\,Jarvis, Y.\,Ruan, The Witten equation, mirror symmetry and quantum singularity theory, arXiv:0712.4021.
\bibitem{for-3}H.\,Fan, T.\,Jarvis, Y.\,Ruan,  The Witten equation and its virtual fundamental cycle, arXiv: math/0712.4025.
\bibitem{fjr-d4} H.\,Fan, T.\,Jarvis, Y.\,Ruan,  Witten's $D_4$ Integrable Hierarchies Conjecture,
arXiv:math/1008.0927.
\bibitem{getzler} E.\,Getzler, The jet-space of a Frobenius manifold and higher-genus Gromov--Witten invariants, Yu.I.\,Manin's Festschrift
\bibitem{givental} A.\,B.\,Givental, Elliptic Gromov--Witten invariants and the generalized mirror conjecture, in Integrable
systems and algebraic geometry, Proceedings of the Taniguichi Symposium 1997, ed. M.\,H.\,Saito, 
Y.\,Shimizu and K.\,Ueno, World Scientific (1998), 107-155.
\bibitem{gm} A.\,B.\,Givental, T.E.\,Milanov, Simple singularities and integrable hierarchies, In: The breadth of symplectic and Poisson geometry, 173-201. Progr. Math. {\bf 232}, Birkh\"auser Boston, Boston, MA, 2005
\bibitem{hert} C.\,Hertling, Frobenius manifolds and moduli spaces for singularities, Cambridge University Press, 2002.
\bibitem{manin} Yu.I.\,Manin, Frobenius Manifolds, Quantum Cohomology, and Moduli Spaces, AMS Colloquium Publications, {\bf 47}, Providence, Rhode Island.
\bibitem{milanov}T. E.\,Milanov, H.-H.\,Tseng, The space of Laurent polynomials, $\mathbb{P}^1$-orbifolds, and integrable hierarchies, J. Reine Angew. Math. vol. 2008, issue 622, p. 189-235.
\bibitem{rossi} P. Rossi, Gromov-Witten theory of orbicurves, the space of
tri-polynomials and Symplectic Field Theory of Seifert fibrations, eprint arXiv:0808.2626v3. 
\bibitem{saito} K.\,Saito, Period mapping associated to a primitive form, Publ. RIMS {\bf 19} (1983) 1231-1264.
\bibitem{stra} I.\,A.\,B.\,Strachan, Symmetries and solutions of Getzler's equation for Coxeter and Extended affine  Weyl Frobenius manifolds, Int. Math. Res. Notices 19 (2003) 1035-1051.
\bibitem{taka} A.\,Takahashi, Weighted projective lines associated to regular systems of weights of dual type, Adv. Stud. Pure Math. {\bf 59} (2010) 371-388.
\bibitem{Witten4} E.\,Witten, Algebraic geometry associated with matrix models of
two-dimensional gravity, In: Topological models in modern mathematics (Stony Brook, NY, 1991), Publish or Perish, Houston, TX, 1993, p. 235-269.
\bibitem{wu} C.-Z.\,Wu, Tau functions and Virasoro symmetries for Drinfeld--Sokolov hierarchies, arXiv:1203.5750.
\end{thebibliography}
\end{document}